\documentclass[journal=iecred,manuscript=article]{achemso}
\usepackage{chemformula} 
\usepackage[T1]{fontenc} 
\usepackage{graphicx}
\usepackage{amsmath}
\usepackage{pdflscape}
\usepackage{multirow}
\usepackage{multicol}
\usepackage{geometry}
\usepackage{textcomp}
\usepackage{float}

\author{Somasekhara Goud Sontti}
\author{Arnab Atta}
\email{arnab@che.iitkgp.ac.in}
\phone{+91 (3222) 283910}
\affiliation[mCFD]
{Multiscale Computational Fluid Dynamics (mCFD) Laboratory, Department of Chemical Engineering, Indian Institute of Technology Kharagpur, Kharagpur, West Bengal 721302, India}
\title[]
{Numerical insights on controlled droplet formation in a microfluidic flow-focusing device}
\begin{document}
\newpage
\begin{abstract}
In this article, we have developed a computational model to determine the droplet formation regime and its transition in a square microfluidic flow-focusing device that eventually dictate the droplet shape, size, and its formation frequency. We have methodically explored the influences of various physico-chemical parameters on the droplet dynamics and flow regime transition, which are essential in the development of new methods for on-demand droplet generation. Based on the droplet formation mechanism, we have formulated flow maps for different liquid-liquid systems, and have also proposed a scaling law to predict the droplet length for a wide range of operating condition resulting from the variation of flow rates, and viscosities of the continuous phase as well as the interfacial tension. This work can effectively contribute in providing helpful guidelines on the design and operations of droplet-based flow-focusing microfluidic systems.
\end{abstract}
\newpage
\section{Introduction}
Microfluidic devices have emerged as a novel tool for conducting chemical and biomedical processes in a miniaturized platform for the past two decades~\cite{ yu2015demand,dang2013reactivity,bennett2018microfluidic,nieves2012hydrodynamics}. Droplet-based microfluidic systems play a significant role in the field of chemical, lab-on-a-chip, drug delivery, biochemical reactions, encapsulation, and nanomaterial synthesis~\cite{song2006,haeberle2007,chueluecha2017enhancement,campbell2018continuous}. Typically in microfluidic devices, reagents volumes can significantly decrease to micro/nanoliters and reaction time could be shortened to a few seconds~\cite{chand2015electroimmobilization,abolhasani2016}. Furthermore, each microfluidic droplet acts as an individual microreactor, isolated from the surrounding phase in which mixing or reaction can occur. Therefore, precise control of droplet size and monodispersity are imperatively demanded in such applications. For the generation of those bubbles/droplets, the most frequently used microfluidic geometries are co-flow, T-junction\cite{ma2018manipulation}, and flow-focusing devices~\cite{teh2008droplet,bordbar2018slug}. Among numerous variants, the flow-focusing device can allow for high-throughput production of monodisperse droplets because of symmetric rupturing of the dispersed phase at the cross junction~\cite{anna2003formation,dang2012preparation}. \citet{tan2008drop} studied the formulation of monodisperse oil-in-water (O/W) and water-in-oil (W/O) emulsions using flow-focusing device, where two-phase flow patterns were investigated for various flow rate ratios, continuous phase viscosities, and surfactant concentrations. \citet{fu-2012s} explored droplet breakup dynamics in viscous liquids and developed flow pattern maps based on dispersed phase Weber number (We) and continuous phase Capillary number (Ca). They observed dripping and jetting flow patterns and proposed a power-law scaling relation for droplet size with respect to the variation of dispersed phase thread. With the help of micro-particle image velocimetry ($\mu$PIV), \citet{ma2014flow} analyzed droplet formation in a flow-focusing microchannel and investigated the flow field inside a droplet. Their findings revealed that droplet topology, internal velocity magnitude, and re-circulation strongly depend on viscosity ratio, $Ca$, and geometric configuration. With increasing viscosity ratio, velocity field inside the droplet changed in the central zone of the droplet. Their results also revealed that by tuning the viscosity ratio, the strength of internal circulatory motions can be controlled. \citet{chen2015model} developed a mathematical model that described the droplet formation in a flow-focusing device under squeezing regime. \citet{kovalchuk2018effect} studied the influence of surfactant concentration on droplet formation in a cross junction microchannel, where various flow regimes such as squeezing, dripping, jetting, and threading were identified. They constructed and compared the flow regime maps for surfactant-free and with-surfactant systems, which showed transition from dripping to jetting and threading at lower flow rates of the dispersed phase in the presence of surfactant. \citet{wu2017liquid} also presented flow regime maps based on the $Ca$ and $We$ for five immiscible liquid-liquid systems in flow-focusing devices with different hydraulic diameters. They described the effect of channel hydraulic diameter, dispersed phase viscosity, and interfacial tension on flow regime transition. Three distinct flow patterns, viz., tubing/threading, dripping, and jetting, were identified. \citet{wu2013ferrofluid} investigated ferrofluid droplet formation in a flow-focusing device by applying radial and axial magnetic fields. Their results provided new insights into the controlling of droplet generation under external force. The applied magnetic field was found to facilitate the formation of longer droplets, but the influence of magnetic field decreased with the increasing flow rates. \citet{varma2017magnetic} reported \textit{Janus} particle formation in the flow-focusing device by applying a magnetic field. \citet{du2016} analyzed the breakup dynamics of the viscoelastic dispersed thread in a flow-focusing device, where four breakup regimes of the viscoelastic thread were observed, and the liquid-liquid interface shape evaluation for each types were presented. Their study provides a foundation for the formation of viscoelastic droplets in microfluidic devices. \citet{sonthalia2016formation} demonstrated the generation of extremely fine water droplets in bitumen solutions using a microfluidic flow-focusing device.

To diminish the necessity of rigorous and exhaustive experimental investigation, a well-validated computational model can aid in exploring critical aspects of physical phenomena that are not attainable by experiments. Furthermore, a predictive model also assists in interpreting the influences of fluid physico-chemical properties on the droplet breakup mechanism, and its plausible shape. For the design of microfluidic devices operating in droplet flow regime, prior knowledge of various parameters such as desired droplet size, shape, velocity, surrounding film thickness, and pressure drop are of paramount importance. Few numerical studies on droplet formation in flow-focusing device were carried out using phase field approach~\cite{zhou2006formation,bai2017three}, Lattice Boltzmann method (LBM)~\cite{wu2008three,gupta2014droplet}, volume of fluid (VOF)~\cite{ong2007experimental,hoang2013benchmark}, and level set (LS)~\cite{li2015control,lan2014cfd} techniques. One of the key steps in such methodologies is the surface tension modeling, and improper model implementation in balance between capillary force and pressure jump across the interface can lead to the evolution of unphysical velocities near the interface, known as \textit{spurious currents}~\cite{popinet1999,harvie2006}. Several researchers have attempted to reduce spurious currents with different approaches~\cite{popinet2003gerris,francois2006balanced,Worner2012,denner2014fully,guo2015}. \citet{sussman2000} developed a coupled LS and VOF (CLSVOF) technique that smoothly captured the continuous interface by calculating the radius of curvature from the LS function. This method has later been successfully utilized to unravel the physics in various applications, e.g., bubble rise in viscous liquids~\cite{keshavarzi2014}, bubble formation on sub-merged orifices~\cite{buwa2007}, droplet impact on a liquid pool~\cite{ray2015}, Taylor bubble formation~\cite{sontti2017numerical,sonttiunderstanding,sontti2018formation}, axisymmetric droplet formation~\cite{chakraborty2016}, droplet coalescence~\cite{mino2016}, and demulsification~\cite{kagawa2014}.
In this work, we present a comprehensive numerical study based on the CLSVOF method, which is arguably better in interface tracking for low Capillary number systems than the conventional VOF technique to delineate the droplet formation mechanism and flow patterns in a microfluidic flow-focusing device. Determining the droplet formation regime and its transition that eventually dictate the droplet shape, size, and formation frequency is the unique contribution of the present study. It is noteworthy that few previous studies have identified the flow regimes and its transition for a limited range of flow rates between 1-600 $\mu$L/min.~\cite{costa2017studies,cao2018dimensionless} We aim to explain the influences of physico-chemical parameters on the droplet dynamics and flow regime transition for a wider range of flow rate (800-2400 $\mu$L/min), which are essential in the development of new high throughput methods for on-demand droplet generation in the areas of drug delivery, crystallization, and materials synthesis. Moreover, a deterministic model for identifying flow regime transition is presented in our work, which to the best our knowledge is missing in the literature. From the perspective of achieving desired droplet size and shape by regulating the flow regime resulting from tuning of operating condition, predictability of the on-demand droplet generation is of immense importance. However, such operation necessitates the understanding of flow regime transition and its manipulation, which is presented here. We systematically explore the effect of flow rates, viscosity ratio, and interfacial tension on droplet length, velocity, volume, and shape. We also offer a unified scaling relation to predict the droplet length for a wider range of operating condition, and develop flow maps for different liquid-liquid systems. Such details in a microfluidic flow-focusing device are absent in the reported literature, which can notably assist in setting guidelines to manipulate droplet size and shape in the flow-focusing devices. 

\section{Problem description}
\begin{figure}[!ht]
	\centering
\includegraphics[width=0.9\linewidth]{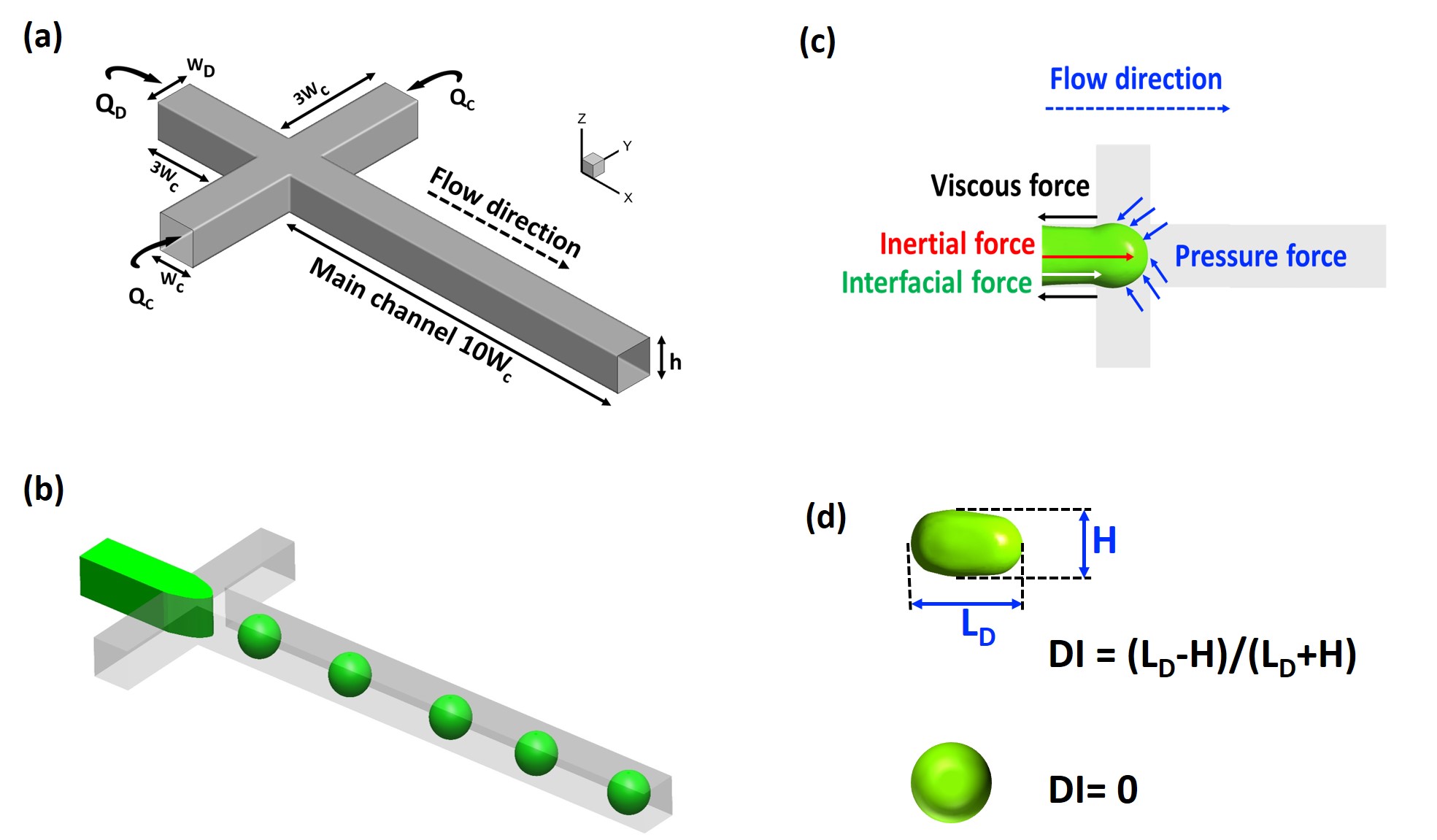}
	\caption{\label{fig:M1}Schematic of (a) computational domain, (b) typical droplet flow in a flow-focusing microchannel, (c) forces acting on the dispersed phase during droplet formation at the cross-junction, and (d) plug and spherical droplets with the definition of droplet deformation index (DI).}
\end{figure}

Schematic of a flow-focusing microchannel considered in this study is illustrated in Fig.~\ref{fig:M1}a, which is of square cross-section having 600~$\mu$m width ($W_c=W_d=W$) $\times$ 600~$\mu$m height ($h$). The lengths of continuous and dispersed phase inlets are 3$W_c$ each, and the length of primary channel is 10$W_c$. The dispersed phase (oil: octane+2.5\% SPAN 80) is introduced through the main channel, and the continuous phase (water/water-glycerol) is entered from the two side channels. 
Typically, in a flow-focusing device, both fluids meet at the cross junction, where the dispersed phase droplet formation occurs due to symmetrical shearing action of continuous phase fluid. Generated droplets then move along the downstream of the microchannel by pushing the continuous phase, as depicted in Fig.~\ref{fig:M1}b. The governing forces such as surface tension, shear stress and pressure build-up at the cross junction arising from the interaction between phases are elaborated in Fig.~\ref{fig:M1}c. The interplay among these forces eventually determines the droplet shape, volume and generation frequency. In this study, the droplet shape is quantified by a deformation index ($DI$), which is based on the length and height of the droplet, as shown in Fig.~\ref{fig:M1}d. It indicates an undeformed/spherical shaped droplet at $DI=0$, and deformed droplets with a length greater than its height are identified by $DI~>~0$.


\section{Numerical method}
Here, we present our CLSVOF model that combines the LS and VOF method. In such a model, the mass and momentum equations are dealt using VOF method, and the fluid interface between two phases is resolved by LS method.  
\subsection{Interface capturing technique} 
In LS method, the fluid phase is identified based on the sign of a marker function ($\varphi$), which depends on position vector ($\vec{\chi}$) and time ($t$). The LS function ($\varphi$) represents the signed distance from the interface, and is calculated as follows~\cite{suss-1994}:  
\begin{equation}
\frac{\partial   \varphi }{\partial t}  +  	\vec{U} .  \nabla  \varphi=0  
\end{equation}
where $\vec{U}$ and $\varphi$ are the velocity and LS function, respectively. The LS function varies smoothly between positive and negative magnitudes of $\textit{d}$, where $\textit{d}=d(\vec{\chi})$ is the shortest distance of a point $\vec{\chi}$ from interface at time $\textit{t}$ and zero value is assigned at the interface. 
\begin{equation}
\varphi ( \vec{\chi },{{t} } ) = \begin{cases}d  & \text{if} \hspace{5px} \text{$\chi$ is  in the liquid phase}\\0 & \text{if} \hspace{5px} \text{$\chi$ is  in the interface}\\-d & \text{if} \hspace{5px} \text{$\chi$ is in the gas phase} \end{cases}  
\end{equation}
In this work, the density and viscosity of each fluid are assumed to be constant, which leads to assigning two different values depending on the sign of LS function in the solution domain. To ascertain continuous variation across the interface, these properties are consequently calculated using a smoothed Heaviside function $\textit{H}(\varphi)$~\cite{sussman2000}. 

\begin{equation}
\rho (\varphi )=\textit{H}(\varphi)\rho_{2}+(1-\textit{H}(\varphi))\rho_{1}  
\end{equation} 
\begin{equation}
\eta(\varphi)=\textit{H}(\varphi)\eta_{2}+(1-\textit{H}(\varphi))\eta_{1}
\end{equation}

The smoothed Heaviside function ($\textit{H}(\varphi)$) is defined as:

\begin{equation}
\textit{H}(\varphi)=\begin{cases} 0& \textit{ if} \hspace{5px} \varphi  < - a  \\  \frac{1}{2}[1+ \frac{ \varphi }{a} + \frac{1}{ \pi }sin( \frac{ \pi  \varphi }{a} ) ]  & \textit{ if} \hspace{5px} |  \varphi  |  \leq a\\ 
1 & \textit{ if} \hspace{5px} \varphi  > a
\end{cases}  
\end{equation}
where \textit{a} is the interface thickness.

\subsection{Fluid flow governing equations}

In CLSVOF approach, a single set of conservation equations is solved for immiscible fluids, as follows~\cite{hirt-1981}.

\textbf{Equation of continuity:} 

\begin{equation}
\label{eq:mass_eqn}
\frac{\partial  \rho }{\partial t}  +  \nabla .  ( \rho  \vec{ U } ) =0
\end{equation}

\textbf{Equation of motion:}

\begin{equation}
\label{eq:mom_eqn}
\frac{ \partial (\rho \vec{ U })}{ \partial t} + \nabla.( \rho \vec{ U } \vec{ U }) = - \nabla P + \nabla.\overline{\overline \tau} + \vec{ F}_{SF}
\end{equation}

\begin{equation}
\label{eq:tau}
\overline{\overline \tau}= \eta \dot{ \gamma } = \eta (\nabla \vec {U} + \nabla { \vec {U} } ^{T})
\end{equation}

where $\vec{U }$, $\rho $, $\eta $, $P$, and $\vec{ F}_{SF}$ are velocity, density, dynamic viscosity of fluid, pressure, and surface tension force, respectively.

%
%

\textbf{Surface tension modeling:}
In CLSVOF, the volumetric surface tension force ($\vec{ F}_{SF} $ in Eq.~\ref{eq:mom_eqn}) is evaluated by modifying the continuum surface force (CSF) model~\cite{brack-1992}, as follows~\cite{sussman1999}:

\begin{equation}
\vec{ F}_{SF} = \sigma  \kappa ( \varphi ) \delta ( \varphi ) \nabla \varphi 
\end{equation}

where $\kappa ( \varphi )$ and $\delta ( \varphi )$  are the interface curvature and the smoothed Dirac delta function, respectively, defined as:

\begin{equation}
\kappa ( \varphi )= \nabla .  \frac{ \nabla  \varphi }{ |  \nabla  \varphi  | } 
\end{equation}
\begin{equation}
\delta ( \varphi )=\begin{cases}0 & \textit{ if} \hspace{5px} |  \varphi  |  \geq a\\ \frac{1}{2a} (1+cos( \frac{ \pi  \varphi }{a} )) & \textit{ if} \hspace{5px} |  \varphi  |  < a\end{cases} 
\end{equation}

\subsection{Solution methodology}
\begin{figure}[!ht]
	\centering
	\includegraphics[width=0.75\linewidth]{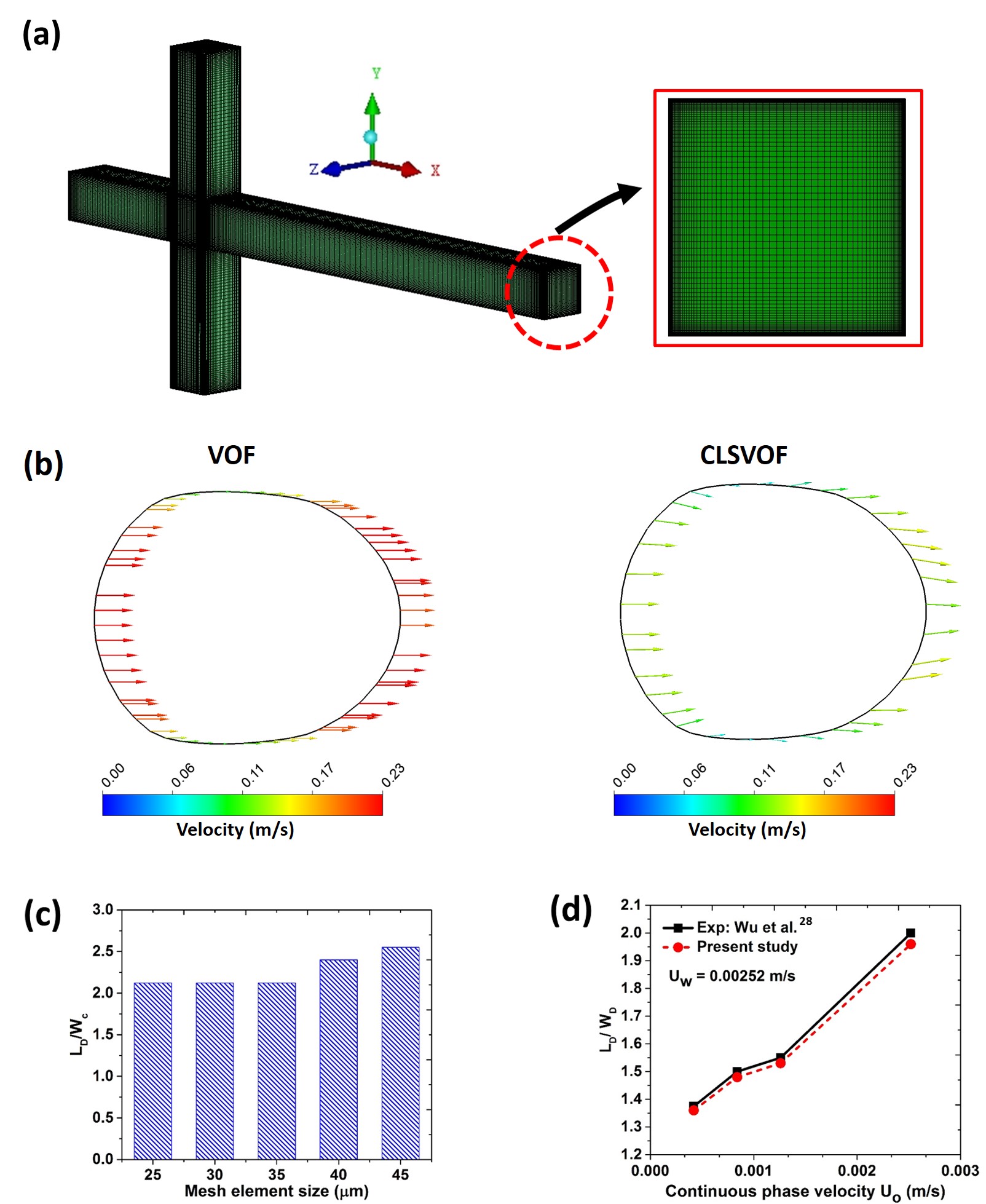}
	\caption{\label{fig:V123} (a) Computational grid with near wall mesh refinement, (b) comparison of parasitic currents around droplet interface through VOF and CLSVOF methods for oil\textendash water system at $Q_w/Q_o=5$ and $Q_o$= 400~$\mu$L/min, (c) grid independence analysis with different mesh element sizes, and (d) comparison of droplet length  predictions against the experimental results of \citet{wu2008three} for $U_w$ =0.00252, $\gamma$ = 30~mN/m, and $\eta_w/\eta_o$ = 0.416.}
\end{figure}

A finite volume method based CFD solver, ANSYS Fluent, is used to solve the aforementioned time-dependent partial differential equations. Pressure implicit with splitting operators (PISO) logarithm is used to solve the pressure\textendash velocity coupling in momentum equation~\cite{issa1986}. The spatial derivatives in momentum and level set equations are discretized using the second-order upwind scheme. The volume fraction is solved using piecewise linear interface construction (PLIC) algorithm~\cite{holt2012}. Subsequently, variable time step and fixed Courant number (Co = 0.25) are considered for simulating the governing equations. In the simulation, constant velocity boundary condition is imposed for both continuous and dispersed phase inlets. At the solid wall, no\textendash slip condition is applied. Pressure boundary is specified at the channel outlet. At the microscale, wall adhesion property plays significant role, where three phases (continuous, dispersed, and solid wall) are in contact. Here, it is assumed that the continuous phase completely wets the channel wall, and the droplet formation occurs at the junction without spreading on the wall. The fluid and material properties considered in this work are taken from the experimental work of \citet{yao2017}, in which the static contact angle of $\theta = 120$\textdegree~is specified at the solid wall. The computational domain is meshed using hexahedral elements, which is further refined near the wall, as shown in Fig.~\ref{fig:V123}a. To capture the droplet surrounding liquid film precisely, at first, its thickness is estimated from analytical correlations\cite{sontti2017cfdsa} for the corresponding fluid properties and operating conditions. Thereafter, near wall mesh refinement is performed, which typically varied in the range of 40~$\mu$m from the wall in all corners. A cross-sectional view of the computational grid in the microchannel is also shown in Fig.~\ref{fig:V123}a. Although the advantage of CLSVOF method over VOF was elaborated in our earlier works with gas-liquid systems~\cite{sontti2017numerical,sonttiunderstanding,sontti2018formation}; in this work, the spurious currents around the droplet is further compared for establishing the efficacy of CLSVOF in liquid-liquid system. With identical element size, the magnitude of velocity fluctuations at the interface is found to be significantly lower in CLSVOF method than the normal VOF method, as shown in Fig.~\ref{fig:V123}b. Prior to the parametric studies, mesh sensitivity analysis on the droplet length is conducted with different element sizes of 25~$\mu$m, 30~$\mu$m, 35~$\mu$m, 40~$\mu$m, and 45~$\mu$m in the core region, as shown in Fig.~\ref{fig:V123}c. Considering both computational accuracy and time, mesh element size of 30~$\mu$m in the core and 6~$\mu$m near the wall are utilized in this study.

\section{Results and discussion}
\subsection{Model validation}
To examine the efficacy of the developed CLSVOF model, it is, at first, validated with the experimental results of \citet{wu2008three} for oil-water system in a flow-focusing device. Numerical simulations were performed with the same fluid properties and channel dimensions, considered by \citet{wu2008three}. 
Fig.~\ref{fig:V123}d demonstrates the quantitative comparison of model prediction for droplet lengths with the experimental observations reported by \citet{wu2008three}. It shows close agreement with a maximum deviation of 7\% from the experimental results that affirms the model efficacy for further investigation. With this validated model, systematic numerical investigations are carried out to elaborate the influence of various physico-chemical parameters on the droplet formation in a flow-focusing device. Five oil-water systems with different glycerol concentrations are considered for which the physical properties are obtained from the experimental results of \citet{yao2017}. It can be clearly seen from Table~\ref{tab:PP1} that on increasing glycerol concentration, the continuous phase viscosity increases from 0.89 to 3.32 $mPa.s$; however, interfacial tension alters negligibly. Therefore, increase in glycerol concentration eventually demonstrates the effect of continuous phase viscosity.
\begin{table}[!ht]
	\centering
	\caption{Physical properties of continuous and dispersed phase at $25^{\circ}C$ used in this study~\cite{yao2017}.}
	\label{tab:PP1}
	\begin{tabular}{@{}llll@{}}
		\hline
		\begin{tabular}[c]{@{}l@{}}Fluid\\ (Glycerol\textendash wt\% Conc.)\end{tabular} & \begin{tabular}[c]{@{}l@{}}Density, \\ $\rho$ ($kg/m^{3}$)\end{tabular} & \begin{tabular}[c]{@{}l@{}}Viscosity \\ $\eta$  (mPa.s)\end{tabular} & \begin{tabular}[c]{@{}l@{}}Interfacial tension with oil\\ $\gamma$ (mN/m)\end{tabular}  \\ \hline
		Water   & 995.6 & 0.89 & 5.37   \\
		5\%~Glycerol   & 1000.8 &0.94& 5.24  \\
		20\%~Glycerol   & 1042.3 & 1.42 & 5.04  \\
		30\%~Glycerol   & 1068.7&2.03 &4.95\\
		40\%~Glycerol   & 1090.4 & 3.32 &5.06  \\
		Octane+2.5\%~SPAN80   & 698 & 0.53 & - \\
		\hline
	\end{tabular}
\end{table}

\subsection{Effect of continuous phase fluid viscosity}
\begin{figure}[!ht]
	\centering
	\includegraphics[width=\linewidth]{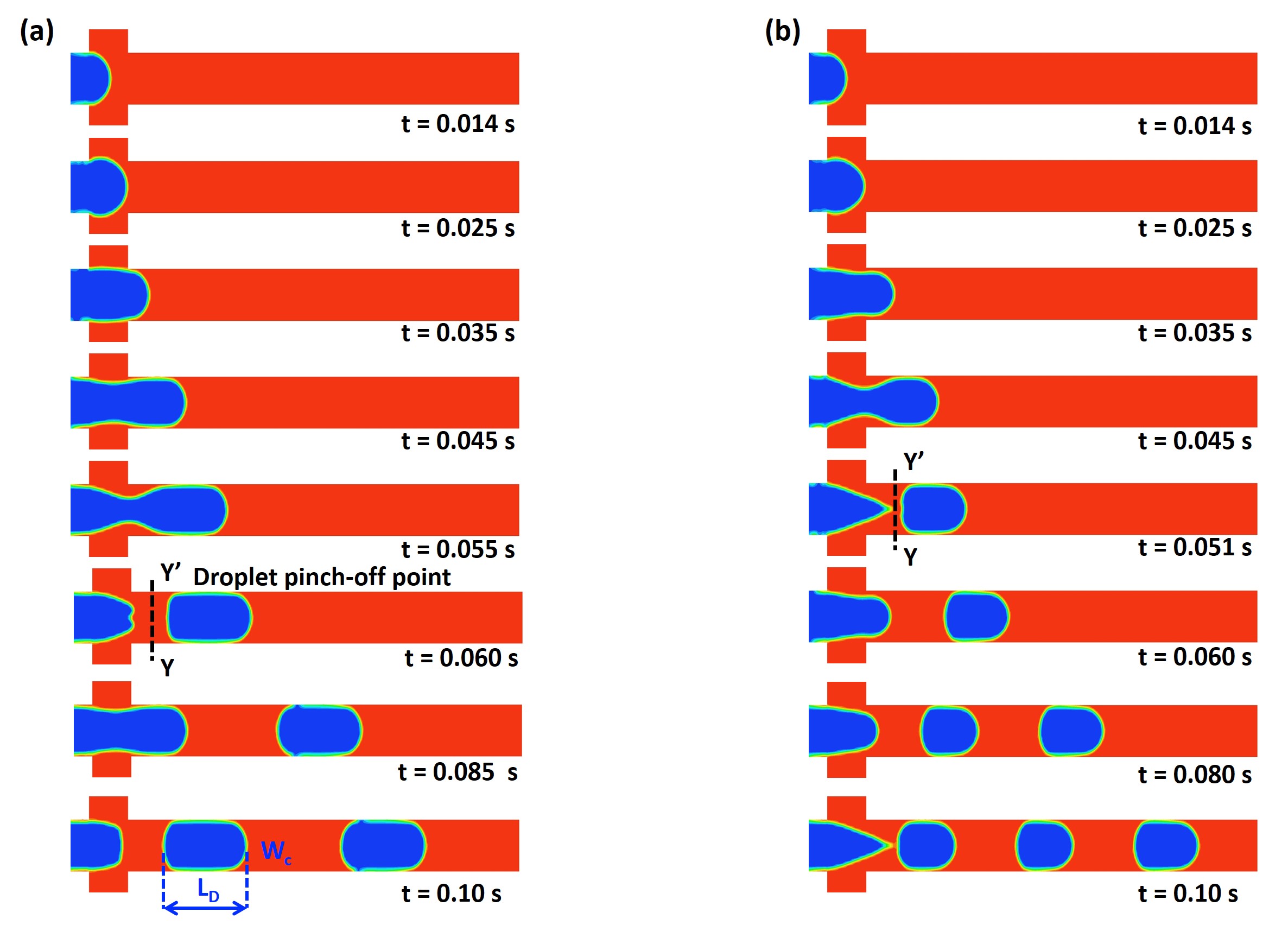}
	\caption{\label{fig:V1}Droplet evolution in a flow-focusing microfluidic device containing (a) oil\textendash water, and (b) oil\textendash water+40\% glycerol at $Q_w/Q_o=2$ and $Q_w$= 400~$\mu$L/min (blue: oil, red: water).}
\end{figure}

In this section, the influence of continuous phase viscosity has been systematically explored on droplet breakup process at a fixed operating condition. Droplet formation in oil\textendash water and oil\textendash water+40\% glycerol systems are illustrated in Fig.~\ref{fig:V1}.
The evolution of dispersed phase growth is evidently influenced by the continuous phase viscosity. In the droplet growth stage, the nose of the droplet partially blocks the outlet channel, which impedes the flow of the continuous phase, as shown in Fig.~\ref{fig:V1}a at $t=0.035~s$. Consequently, droplet head expands slowly in the downstream and the dispersed phase neck is formed. At the cross junction, continuous phase squeezes the dispersed phase symmetrically and pushes the droplet thus formed in the radial and axial directions. Fig.~\ref{fig:V1}a also shows the droplet pinch-off at $t$= 0.060 $s$. At the same operating condition for oil\textendash water+40\% glycerol system, droplet breakup process is similar to the oil-water system, but an early droplet breakup at $t= 0.051~s$ is observed (Fig.~\ref{fig:V1}b). This is ascribed to the increased viscous force that resists the growth of dispersed phase in the main channel. Therefore, droplet formation time decreases with increase in the continuous phase viscosity. Additionally, it is apparent that droplet pinch-off position shifts with fluid properties. Due to higher viscous resistance in the continuous phase, dispersed phase thread length increases, and droplet pinch-off occurs toward the downstream of the main channel in case of oil\textendash water+40\% glycerol system, unlike the oil\textendash water system, where the droplet detaches close to the cross-junction. Typically, the pinch-off position can also be influenced by the operating condition and geometric configuration.

\begin{figure}[!ht]
	\centering
	\includegraphics[width=0.8\linewidth]{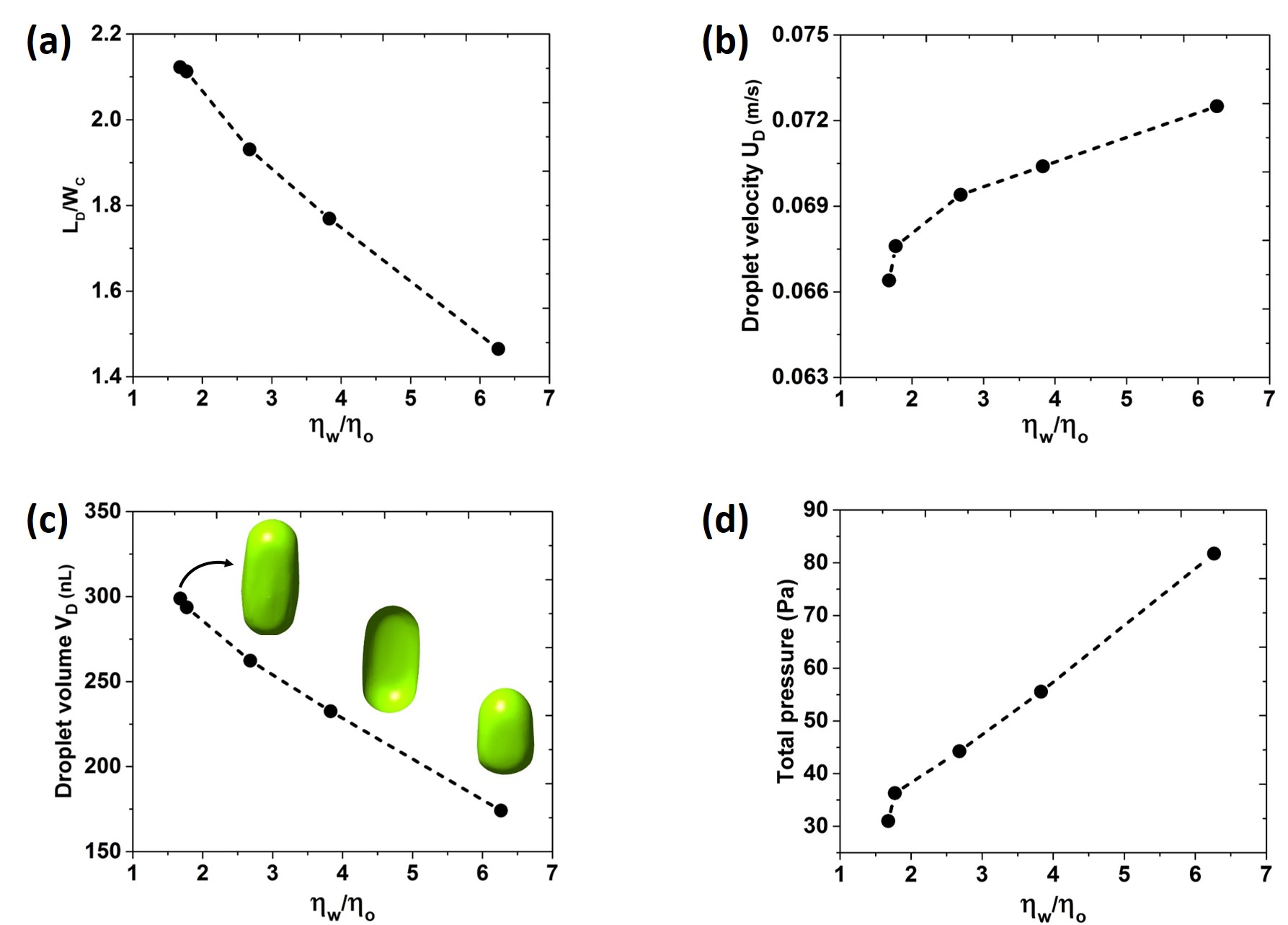}
	\caption{\label{fig:VA1} Effect of continuous phase fluid viscosity on (a) non-dimensional droplet length, (b) droplet velocity, (c) droplet volume, and (d) pressure drop at $Q_w/Q_o=2$ and $Q_w$= 400~$\mu$L/min.}
\end{figure}

\begin{figure}[!ht]
	\centering
	\includegraphics[width=0.7\linewidth]{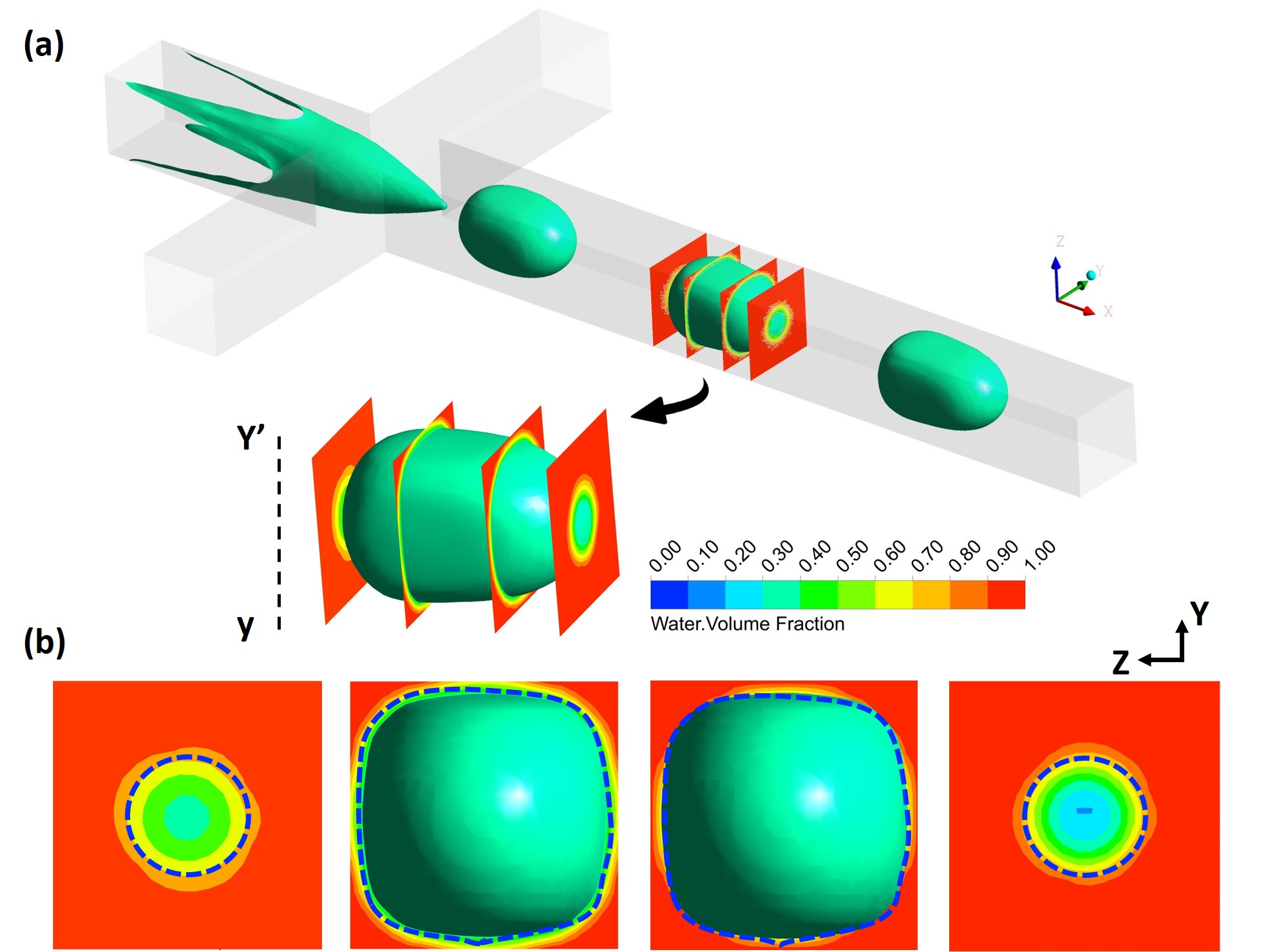}
	\caption{\label{fig:V112} (a) 3-D iso-surface view of simulated droplet formation in the dripping regime, where the droplet along with four vertical slices are shown in zoomed view, and (b) visualization of surrounding liquid film thickness around the droplet (left to right: rear to nose) for $Q_w/Q_o=2$ and $Q_w$= 400~$\mu$L/min.}
\end{figure}

\indent Fig.~\ref{fig:VA1}a demonstrates the effect of continuous phase viscosity on non-dimensional droplet length with respect to channel width.
It is apparent from Fig.~\ref{fig:VA1}a that with increasing continuous phase viscosity, droplet length decreases but its velocity increases, as shown in Fig.~\ref{fig:VA1}b. This can be attributed to increased liquid film thickness around the droplet and its nonuniform shape i.e., different radii of curvatures at the nose and rear, which results in pressure difference around the droplet. The developed pressure gradient leads to increase in droplet velocity with increasing continuous phase viscosity. The volume of individual droplet is also calculated, which significantly decreases with increasing viscosity ratio, as depicted in Fig.~\ref{fig:VA1}c. It can be also noted from Fig.~\ref{fig:VA1}c insets, droplet nose curvatures decreases with increasing continuous phase viscosity. Fig.~\ref{fig:VA1}d shows the effect of viscosity ratio on the total pressure drop, where an increase in viscosity ratio results in almost linear enhancement of total pressure drop. It is well known that the thin film between the channel wall and the droplet is favorable for heat and mass transfer process. This liquid film thickness depends on a number of parameters, including interfacial tension, viscosity ratio of the two flowing phases, and flow rates. Furthermore, it is non\textendash uniform, which varies in both lateral and axial directions from the droplet nose to its tail. Fig.~\ref{fig:V112}a illustrates the three-dimensional iso-surface visualization in dripping flow regime.
To visualize the liquid film thickness around the droplet, four slices are created along $YZ$ plane between nose to tail of the droplet, which are shown in subset of Fig.~\ref{fig:V112}a (in zoomed view of a droplet with slices). Fig.~\ref{fig:V112}b shows the 2D planar views of those four planes. These results further establish the efficacy of CLSVOF method in precise capture of the liquid film thickness around the droplet.
\begin{figure}[!ht]
	\centering
	\includegraphics[width=\linewidth]{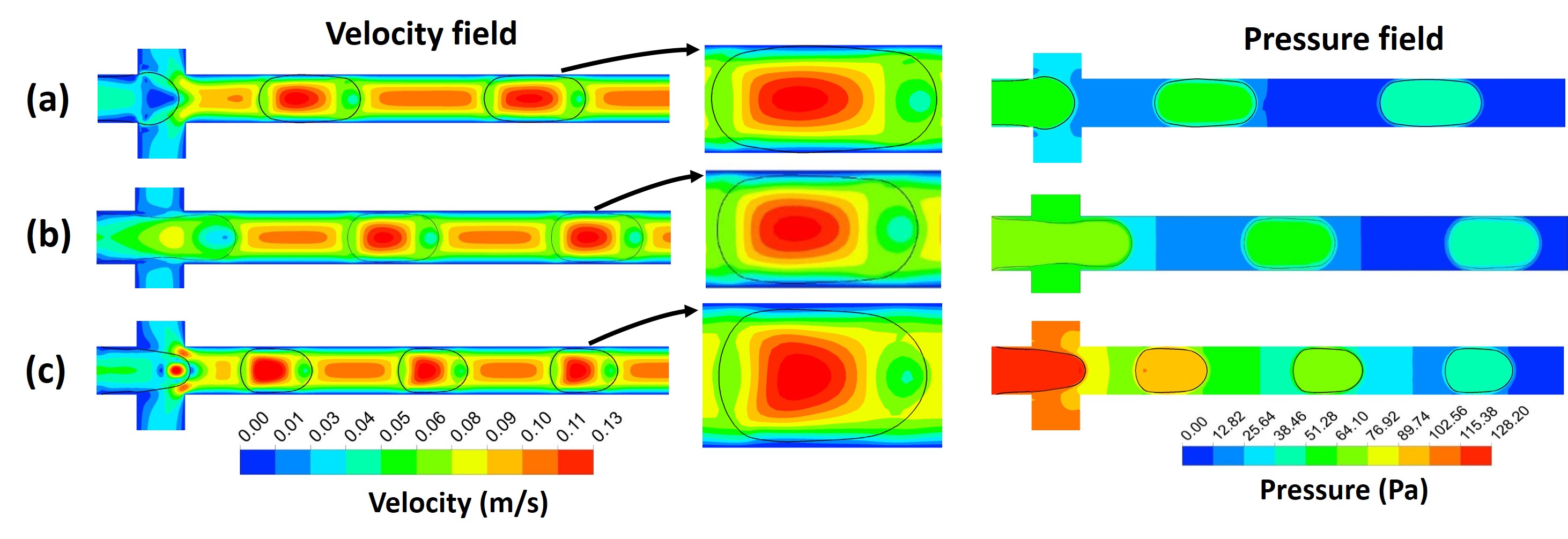}
	\caption{\label{fig:VA7} Effect of continuous phase fluid viscosity on velocity and pressure field distributions in the microchannel for (a) oil\textendash water, (b) oil\textendash water+20\% glycerol, and (c) oil\textendash water+40\% glycerol at $Q_w/Q_o=2$ and $Q_w$= 400~$\mu$L/min. }
\end{figure}

Fig.~\ref{fig:VA7} demonstrates the influence of viscosity on velocity field and pressure distributions in the microchannel. It is found that with an increase in continuous phase liquid viscosity, velocity magnitude inside the droplet increases. This is mainly ascribed to the combined effect of enhanced liquid film thickness, change in droplet shape and viscous stress on the dispersed droplet. Interestingly, velocity magnitude at the nose of a droplet is found to be significantly lower than that of the middle in all cases. This observation is in close agreement with the experimental results of \citet{li2017experimental} identified using $\mu$-PIV. Similar trend is also observed in the liquid slugs. Furthermore, corresponding pressure fields are also analyzed, as shown in Fig.~\ref{fig:VA7} for three different systems with increasing continuous phase liquid phase viscosity. It is evident from the results that pressure inside the droplet and liquid slug also increases with increasing continuous phase liquid viscosity.

\subsection{Effect of interfacial tension}
\begin{figure}[!ht]
	\centering
	\includegraphics[width=\linewidth]{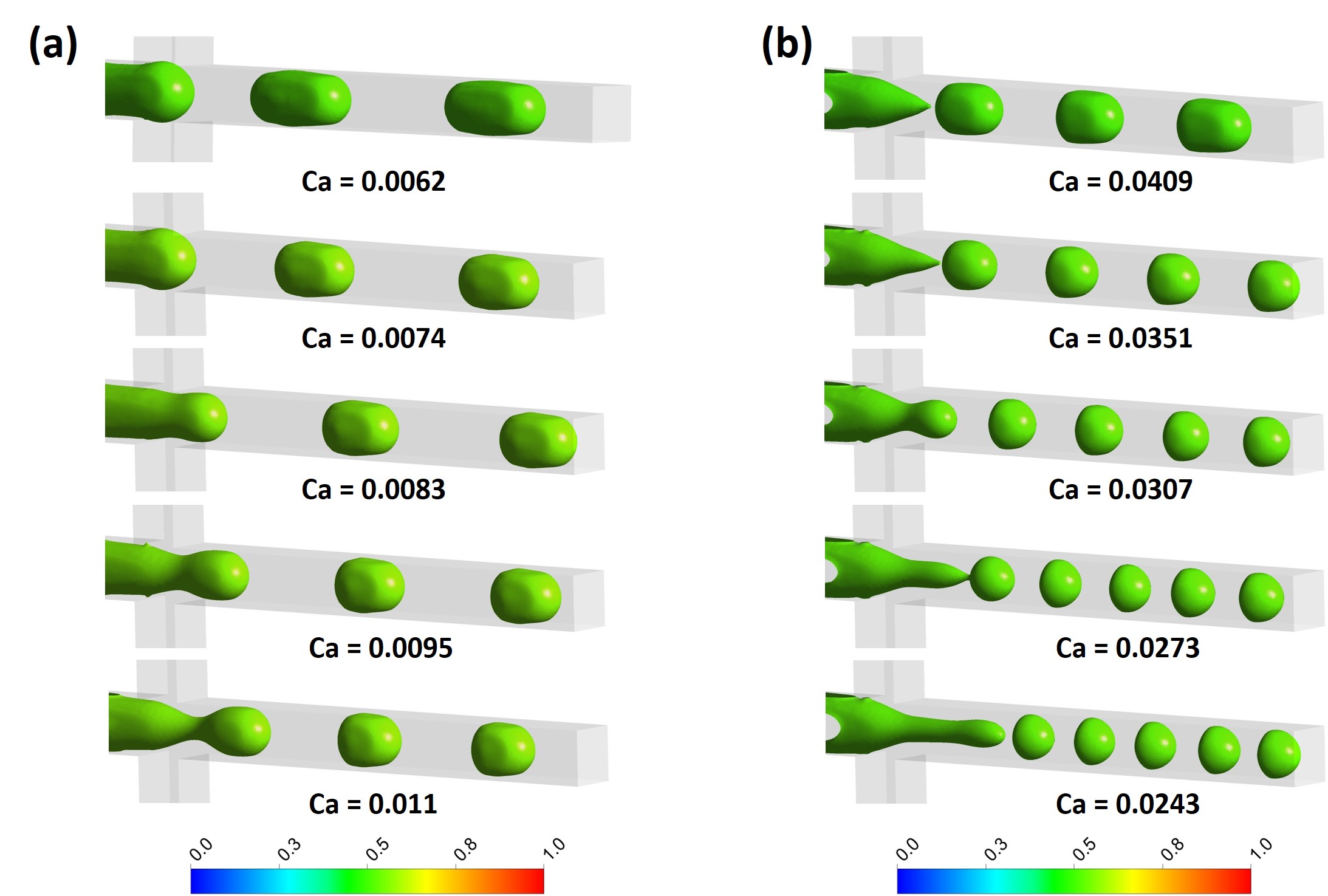}
	\caption{\label{fig:S2} Effect of interfacial tension on droplet shape and flow regime iso-surface visualization for (a) oil\textendash water, and (b) oil\textendash water+40\% glycerol system at $Q_w/Q_o=2$ and $Q_w$= 400~$\mu$L/min.}
\end{figure}

To understand the effect of interfacial tension on droplet formation and flow regime, a set of simulations is carried out keeping other properties unaltered. As disused in the Fig.\ref{fig:M1}c,  the droplet formation in microfluidic devices has a strong dependence on interfacial tension and viscous forces. Typically, the interfacial tension of solution decreases with increasing the surfactants concentration. In this work, the interfacial tension for three different liquid-liquid systems is varied from their respective reference values 5 $mN/m$  to 3 $mN/m$, which are in the practical range \cite{yao2017} that make our results more relevant for application. Subsequently, the results are quantified based on the Capillary number ($Ca=\eta U_{c}/\gamma$).
\begin{figure}[!ht]
	\centering
	\includegraphics[width=\linewidth]{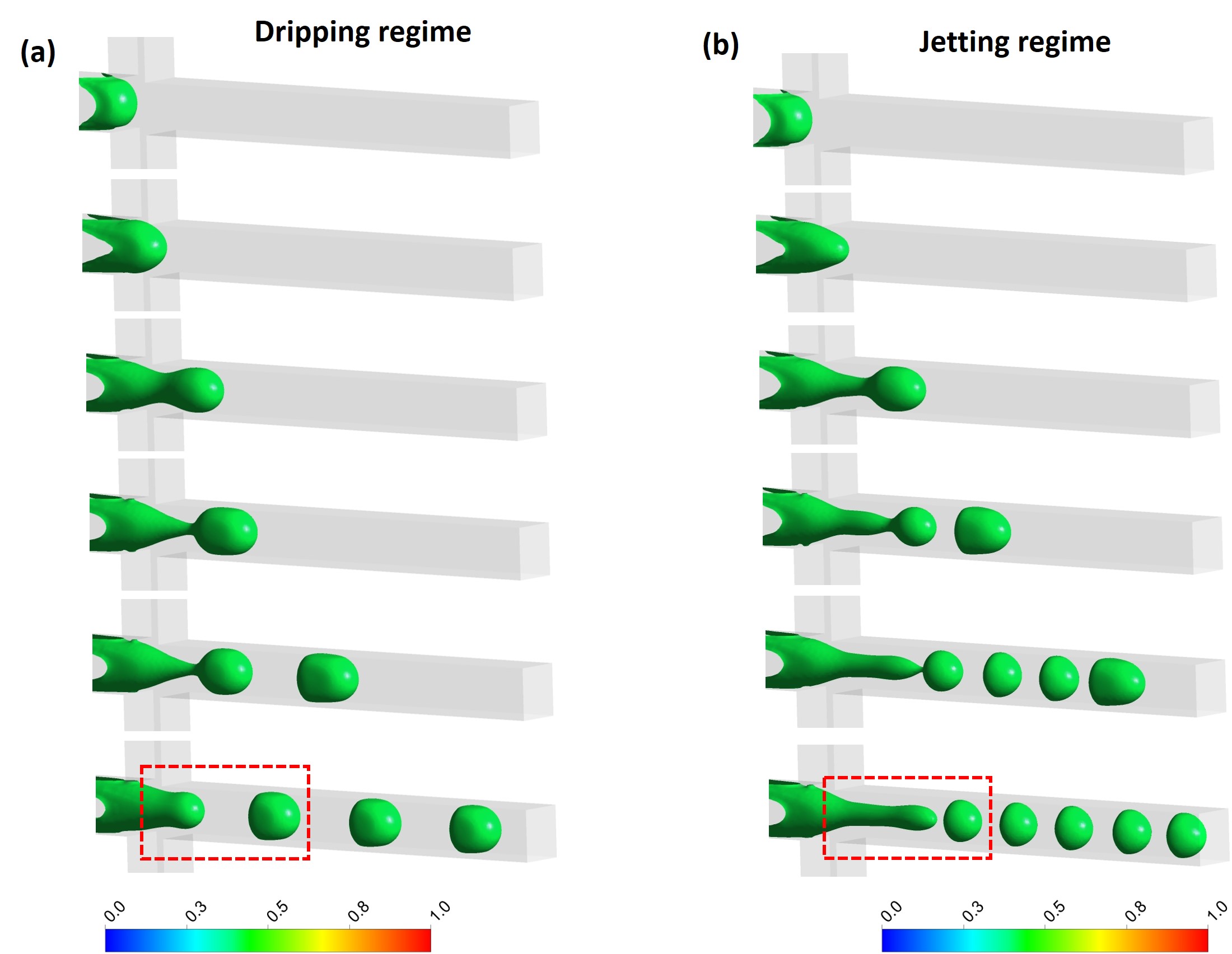}
	\caption{\label{fig:S3}Sequence of images displaying formation of a droplet in oil\textendash water+40\% glycerol system (a) dripping regime at $Ca$ = 0.025, and (b) jetting regime at $Ca$ = 0.041, for $Q_w/Q_o=2$ and $Q_w$= 400~$\mu$L/min.}
\end{figure}
Fig.~\ref{fig:S3} demonstrates the effect of interfacial tension on flow patterns for oil\textendash water+40\% glycerol system. Based on the dominant forces in the flow-focusing device, the formation can be generally classified into the squeezing, dripping and jetting regime. Fig.~\ref{fig:S3}a shows the sequence of three-dimensional iso-surface visualizations in the dripping regime at Ca = 0.025 for relatively low flow rate. However, decrease in the interfacial tension results in shifting of the flow regime from dripping to jetting, as depicted in Fig.~\ref{fig:S3}b. In the jetting regime, a long thread of the dispersed phase is formed and propagated into the downstream of the microchannel, which reaches an approximately constant value (Fig.~\ref{fig:S3}b marked by red color box). Subsequently, the jet propagates fluctuations induced by the Rayleigh-Plateau instability. Interestingly, in this regime the generated droplet diameter is larger than the dispersed thread diameter. 
\begin{figure}[!ht]
	\centering
	\includegraphics[width=\linewidth]{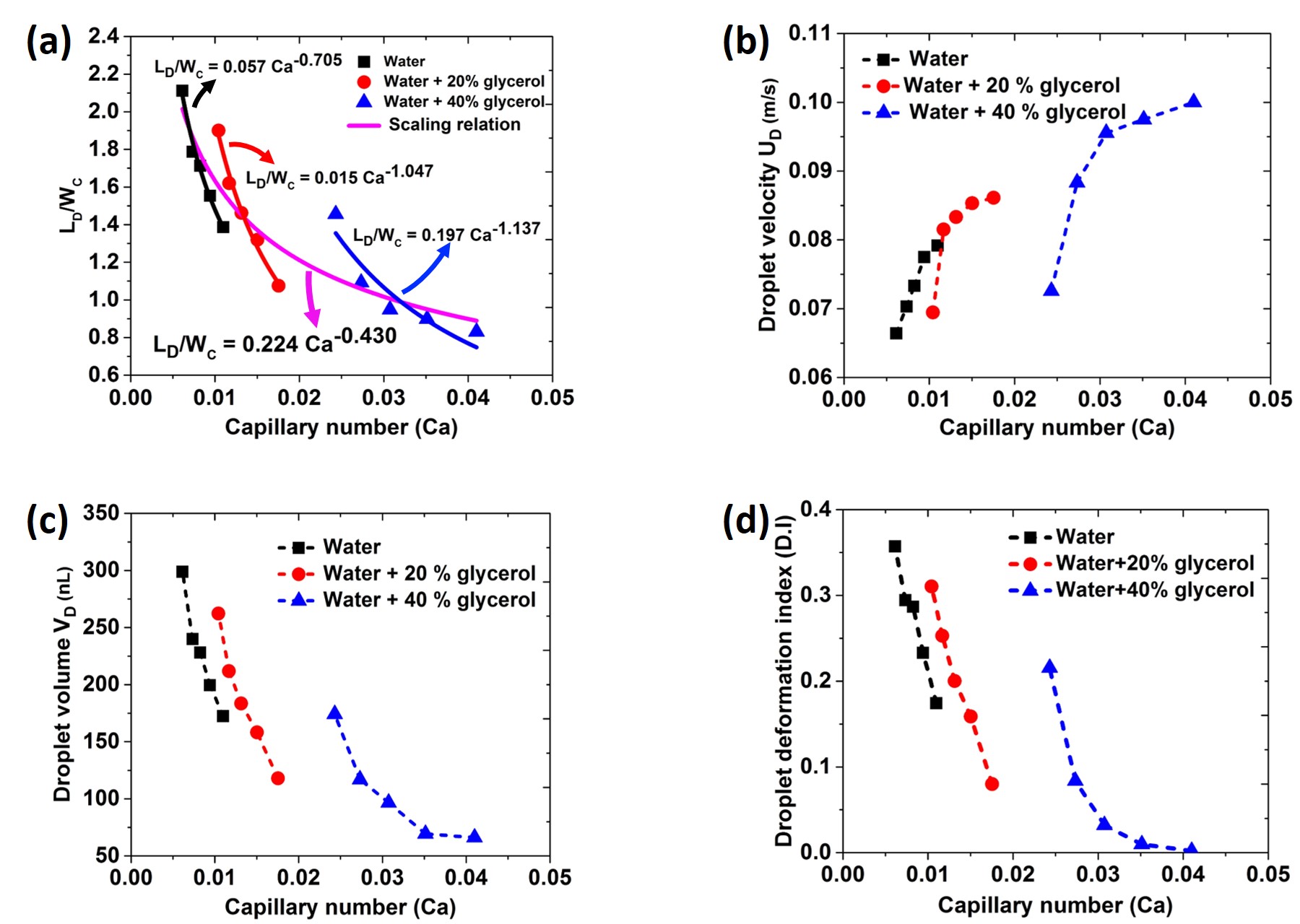}
	\caption{\label{fig:S1} Effect of interfacial tension on (a) non-dimensional droplet length, (b) droplet velocity, (c) droplet volume, and (d)  deformation index (DI) at $Q_w/Q_o=2$ and $Q_w$= 400~$\mu$L/min.}
\end{figure}
Fig.~\ref{fig:S1} shows the effect of interfacial tension on non-dimensional droplet length, velocity, volume, and deformation index for three different oil-water systems. In all cases, droplet length is found to decrease with increasing $Ca$, as shown in Fig.~\ref{fig:S1}a. This is attributed to the weak cohesive forces among liquid molecules. Furthermore, to predict the droplet length, simple power-law scaling relations are proposed as a function of $Ca$ in the considered operating range. These relations are similar to the previously reported results for various gas-liquid and liquid-liquid systems but with different pre-factor and exponent~\cite{sontti2019cfdf,sontti2017cfd,wu2017liquid}. Additionally, a unified scaling relation ($L_{D}/W_{c} = 0.244~Ca^{-0.430}$) is formulated to encompass the studied range of $Ca$ resulting from the variation of interfacial tension, and it shows sound agreement with a maximum deviation of 24\%. Droplet velocity is found to increase with increasing $Ca$ (i.e decreasing interfacial tension) for all cases, as shown in Fig.~\ref{fig:S1}b. As mentioned earlier, this phenomenon can be attributed to increased liquid film thickness and change in droplet shape from elongated plug to near spherical shape. Fig.~\ref{fig:S1}c shows the change in droplet volume with $Ca$ for three different systems. Similar to the droplet length, droplet volume also decreases with increasing $Ca$. Furthermore, droplet shape is quantified by droplet deformation index ($DI$) based on its width and height (Fig.~\ref{fig:M1}d). It can be seen from Fig.~\ref{fig:S1}d that the droplet sphericity increases with increasing $Ca$. This is further elaborated in Fig.~\ref{fig:S2}, which demonstrates the 3D iso-surface view of droplet shapes in microchannel for oil\textendash water and oil\textendash water+40\% glycerol systems.   
It can be inferred from Fig.~\ref{fig:S2}a that with decreasing interfacial tension, droplet length decreases and the shape changes from elongated plugs to smaller plugs. For the studied range of $Ca$, dripping regime is evident from Fig.~\ref{fig:S2}a for oil\textendash water. However, for oil\textendash water+40\% glycerol system, two different flow regimes are observed with increasing $Ca$ as shown in Fig.~\ref{fig:S2}b. In the range of $0.0243\leq Ca \leq0.0307$, dripping regime is noticed, and for $0.0307 < Ca\le 0.0351$, it changes from dripping to jetting. Monodispersity of droplets is one of the important properties in droplet generation. Interestingly, a significant change in droplet shape is also observed for oil\textendash water+40\% glycerol. Droplet shape is changed from plug to near spherical with increasing $Ca$. 
\subsection{Effect of dispersed phase viscosity}
\begin{figure}[!ht]
	\centering
	\includegraphics[width=\linewidth]{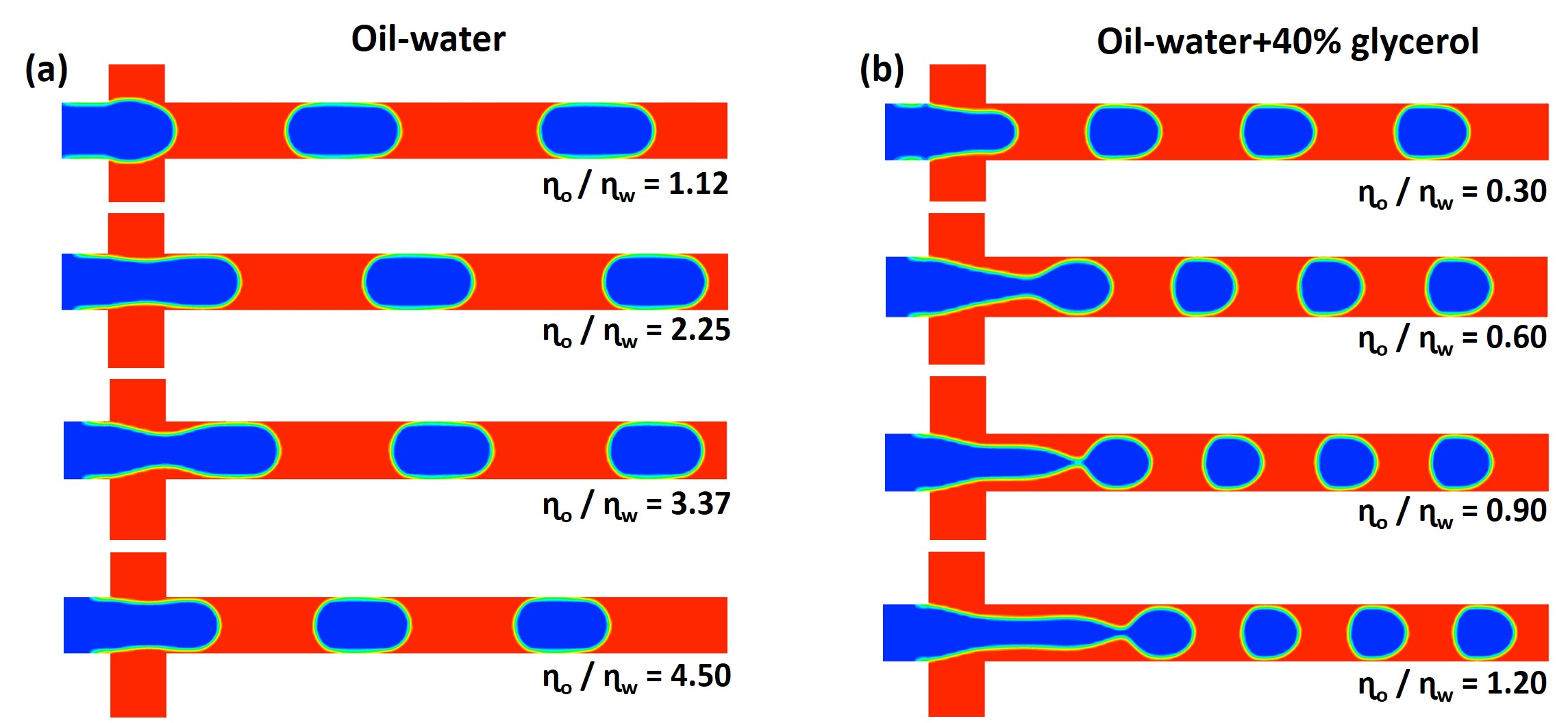}
	\caption{\label{fig:o17}Effect of dispersed  phase viscosity on volume fractions for (a) oil-water, and (b) oil\textendash water+40\% glycerol system at $Q_w/Q_o=2$ and $Q_w$= 400~$\mu$L/min.}
\end{figure}

This section systematically explores the effect of dispersed phase viscosity on the droplet length and velocity by altering the oil viscosity ($\eta_o$= 0.53\textendash4 mPa.s) and keeping other properties and operating conditions constant. 
\begin{figure}[!ht]
	\centering
	\includegraphics[width=\linewidth]{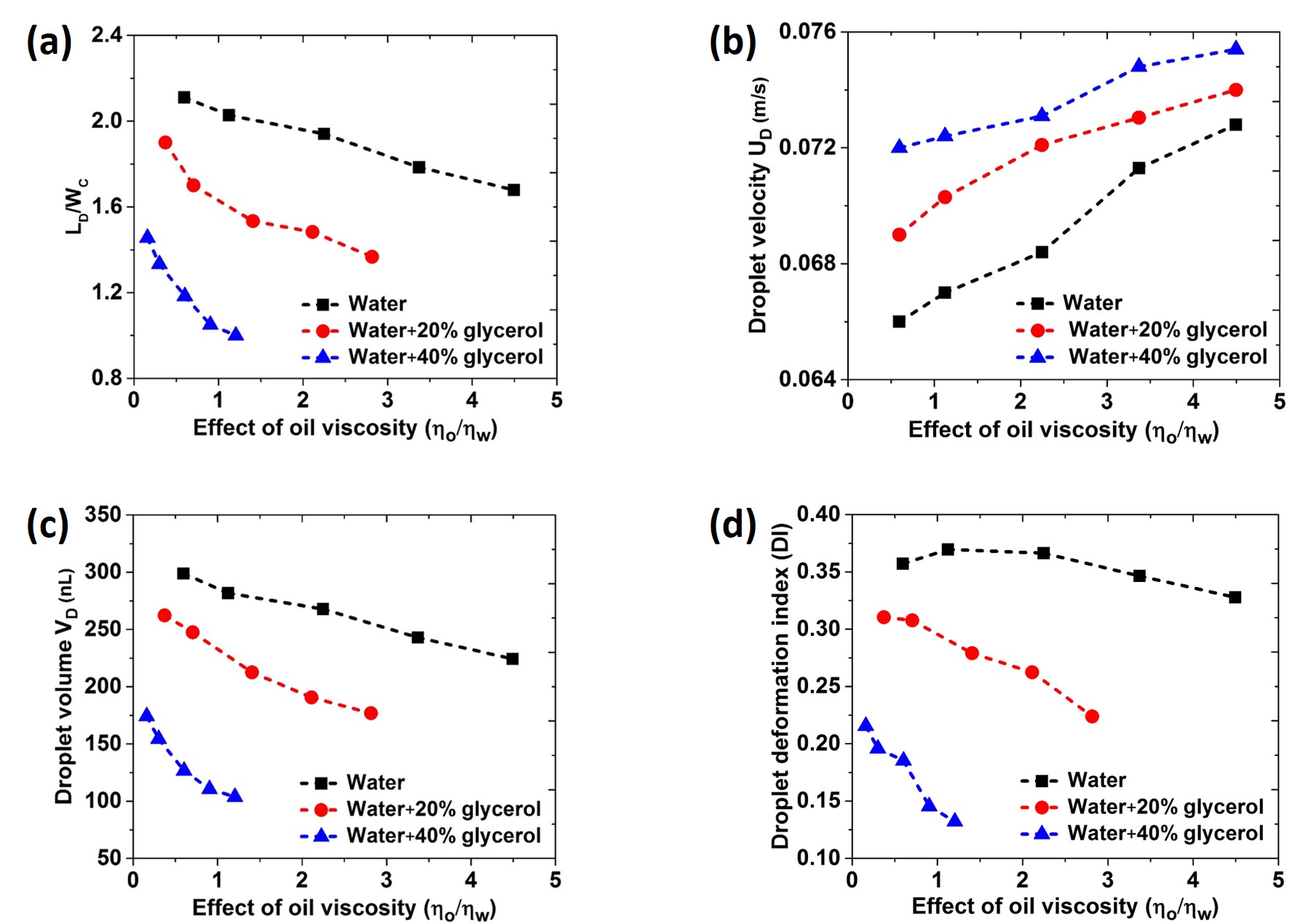}
	\caption{\label{fig:o1}Effect of dispersed  phase viscosity (a) non-dimensional droplet length, (b) droplet volume, (c) droplet volume, and (d) deformation index (DI) at $Q_w/Q_o=2$ and $Q_w$= 400~$\mu$L/min.}
\end{figure}
Fig.~\ref{fig:o1}a shows the influence of oil phase viscosity on droplet length and it indicates that with an increase in oil viscosity, the droplet length decreases for all three systems. This can mainly be attributed to the increased viscous resistance of dispersed phase in the main channel. Consequently, a greater shear force acts on the dispersed phase, which tends to produce smaller droplets at the cross-junction. A significant difference in droplet length is observed for a higher viscous system of oil\textendash water+40\% glycerol as compared to the oil-water system due to higher viscosity of the continuous phase liquid. It is worth noting that oil\textendash water system has relatively higher interfacial tension than oil\textendash water+40\% glycerol. Therefore, the droplets would be longer for oil\textendash water system as compared to oil\textendash water+40\% glycerol, which is discussed in the previous section. This difference in droplet length is attributed to the complex effect of liquid properties such as dispersed phase viscosity and interfacial tension. In line with the previous discussion, droplet velocity is found to increase as a result of different droplet shape and enhancement in associated near wall liquid film thickness with increasing dispersed phase viscosity, as shown in Fig.~\ref{fig:o1}b. Droplet volume is also quantified and is found to decrease with increasing dispersed phase viscosity for all the cases, as shown in Fig.~\ref{fig:o1}c. The $DI$ also decreases (Fig.~\ref{fig:o1}d) as the viscosity ratio increases. It is also found that for lower viscous cases, longer droplets are formed, where the length is greater than channel hydraulic diameter. Therefore, the droplet shape is also found to be influenced by the dispersed phase viscosity. It can be seen from the volume fraction contours (Fig.~\ref{fig:o17}a) that the droplet length decreases with increasing viscosity ratio ($\eta_o/\eta_w=1.12-4.50$) for oil-water system.
In the considered viscosity range, dripping regime is experienced and plug shape droplets are observed. Similarly, for oil\textendash water+40\% glycerol a significant change in droplet length is also observed, as shown in Fig.~\ref{fig:o17}b. For oil\textendash water+40\% glycerol system, dripping regime is observed for lower viscosity range ($\eta_o/\eta_w=0.30-0.60$). With further increase in viscosity ratio, flow regime shifts from dripping to jetting. Droplet shape also changes from smaller plugs to near spherical with increasing viscosity ratio (Fig.~\ref{fig:o17}b).  

\subsection{Effect of flow rate}
\begin{figure}[!ht]
	\centering
	\includegraphics[width=\linewidth]{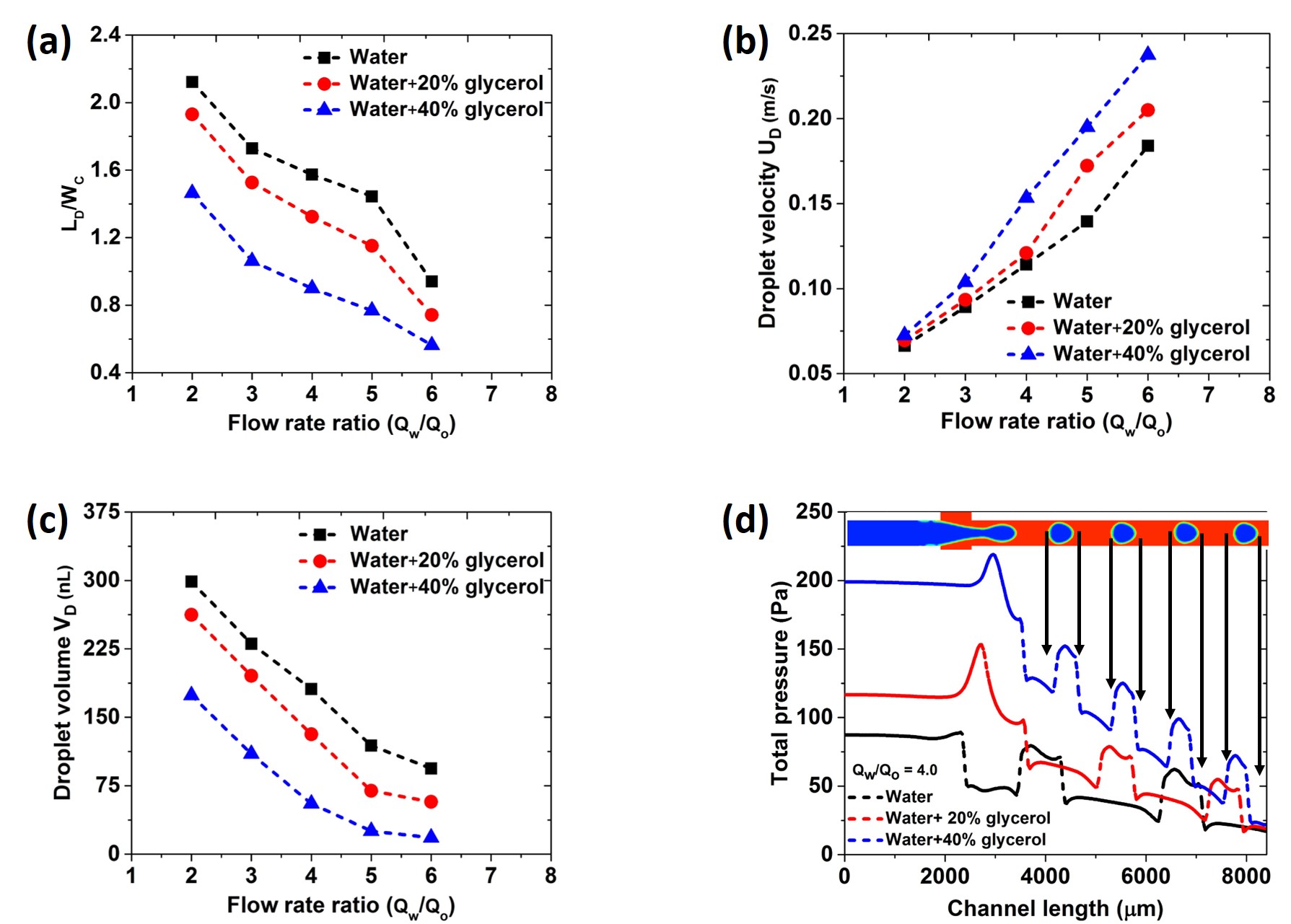}
	\caption{\label{fig:V2} Effect of continuous phase flow rate (a) non-dimensional droplet length, (b) droplet velocity, (c) droplet volume, and (d) pressure profiles along the channel centerline at $Q_o$= 400~$\mu$L/min.  }
\end{figure}

To demonstrate the influence of continuous phase velocity on droplet breakup, systematic investigations are performed by altering the continuous phase flow rate for three different systems. It can be seen from Fig.~\ref{fig:V2}a that the droplet length decreases with increasing continuous phase flow rate at fixed $Q_{o}$. 
This is due to the fact that shear force increases on the dispersed phase at cross junction with an increase in continuous phase flow rate. Consequently, it leads to the formation of smaller droplets with higher formation frequency. It is apparent from Fig.~\ref{fig:V2}b that the droplet velocity increases with increasing the continuous phase flow rate due to increase in liquid film thickness and change in droplet shape. 
Similar to droplet length, droplet volume also decreases with increasing continuous phase flow rate for all cases, as shown in Fig.~\ref{fig:V2}c. The pressure variation along the centre line of the channel is presented in Fig.~\ref{fig:V2}d, which shows that pressure drop increases with increasing continuous phase viscosity (i.e glycerol concentration) at a fixed flow rate. The pressure field in dispersed phase is found to be nearly constant up to the droplet pinch-off point. Thereafter, significant change in total pressure is observed due to the interaction of two-phases. In any pressure profile, each peak and its width indicate a droplet position and shape, respectively. Moreover, due to difference in radius of curvature from rear to nose of a droplet, there exists a pressure gradient across the droplet, which is linked with the pressure profile in the continuous phase. These results corroborate with the theoretical analysis by \citet{abiev2017analysis}, which ensures the link between pressure profile inside and outside of a droplet through Laplace equation.

\begin{figure}[!ht]
	\centering
	\includegraphics[width=\linewidth]{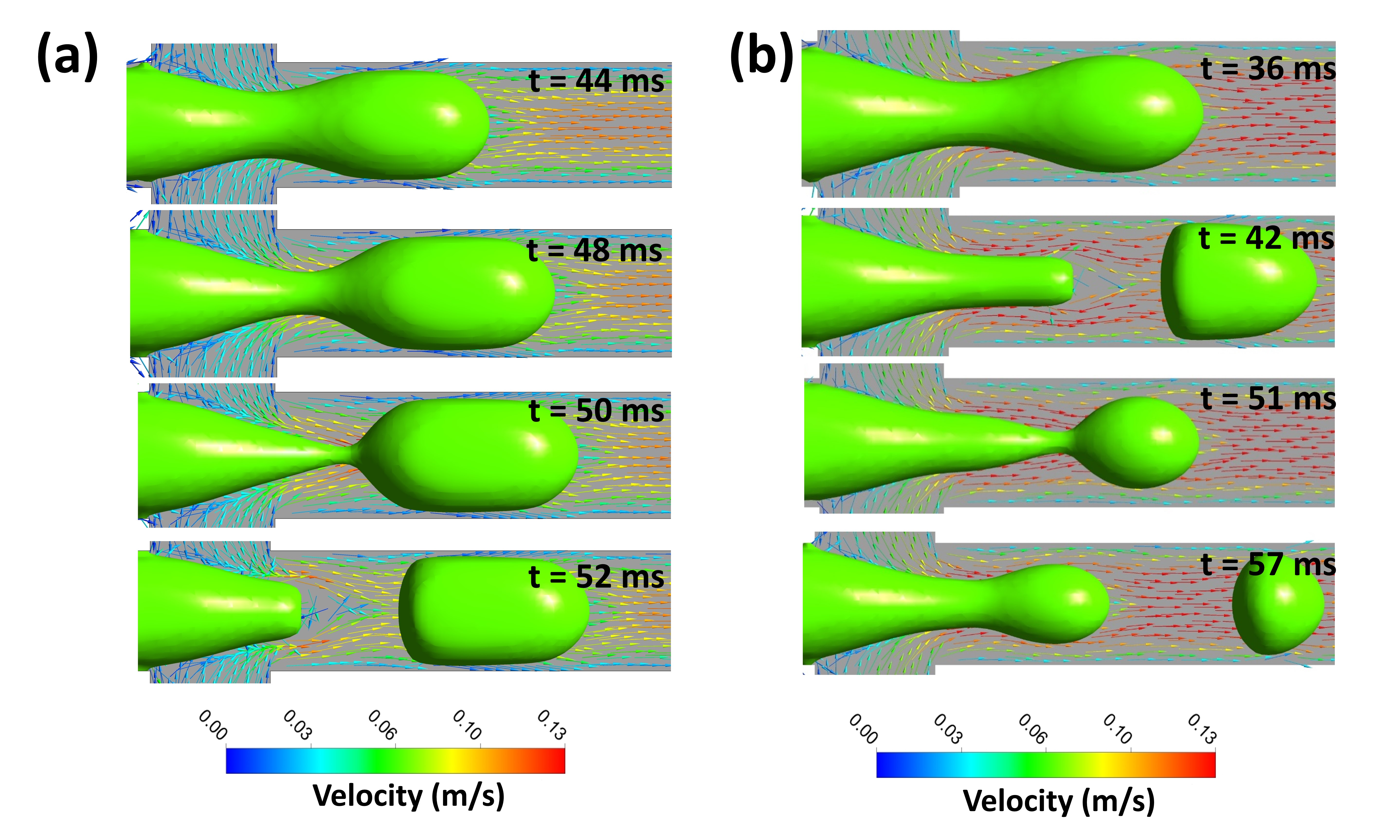}
	\caption{\label{fig:V1245} Comparison of velocity fields during droplet formation in oil\textendash water+40\% glycerol system for (a) $Q_w/Q_o=2$, and (b) $Q_w/Q_o=4$ at $Q_w$= 400~$\mu$L/min.}
\end{figure}

Fig.~\ref{fig:V1245} demonstrates the shearing effect by continuous phase for two different flow rate ratios at fixed $Q_{w}$. 
It can be seen from Fig.~\ref{fig:V1245} that the continuous phase flow rate plays an important role in symmetrical shearing of dispersed phase and also influences the droplet size, formation time, and flow regime. For lower flow rate, first droplet breakup is observed at $t = 51~ms$. However, in case of higher flow rate, it is noticed at $ t= 35~ms$. Velocity field around the dispersed phase is also compared for between those two cases in Fig.~\ref{fig:V1245}, which shows increased velocity magnitude in higher continuous phase flow rate as compared to the lower flow rate.
\begin{figure}[!ht]
	\centering
	\includegraphics[width=0.7\linewidth]{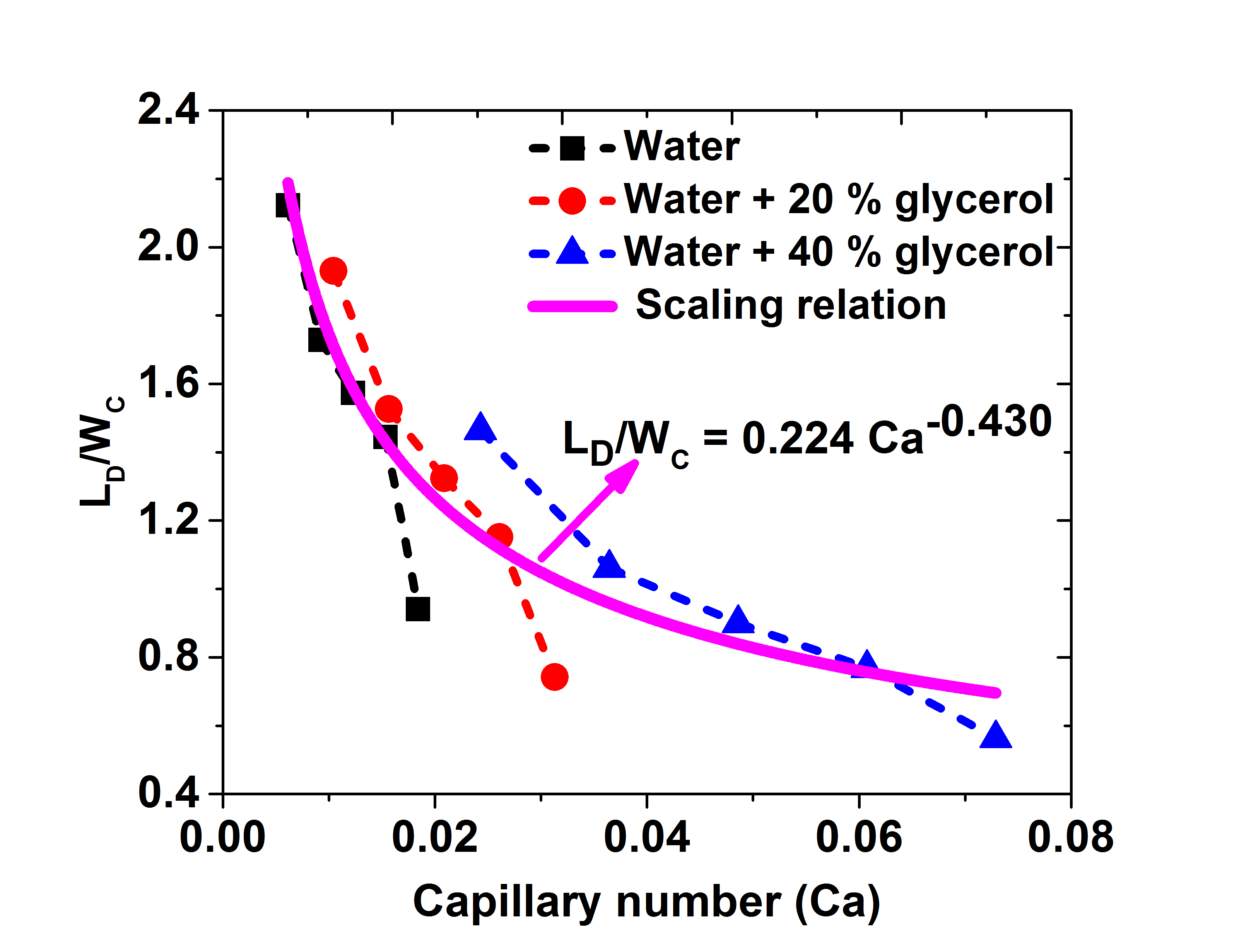}
	\caption{\label{fig:FV2} The scaling relation for non-dimensional droplet length with the Capillary number for different continuous phase flow rates at $Q_o$= 400~$\mu$L/min.  }
\end{figure}
Additionally, the developed scaling relation ($L_{D}/W_{c} = 0.244~Ca^{-0.430}$ in Fig.~\ref{fig:S1}a) is further verified for variation in continuous phase flow rate, as shown in Fig.~\ref{fig:FV2}. The proposed unified scaling relation conforms well with a maximum deviation of 27\% for the complete range of our studied operating condition encompassing the influence of flow rates and viscosities of the continuous phase.       
\begin{figure}[!ht]
	\centering
	\includegraphics[width=\linewidth]{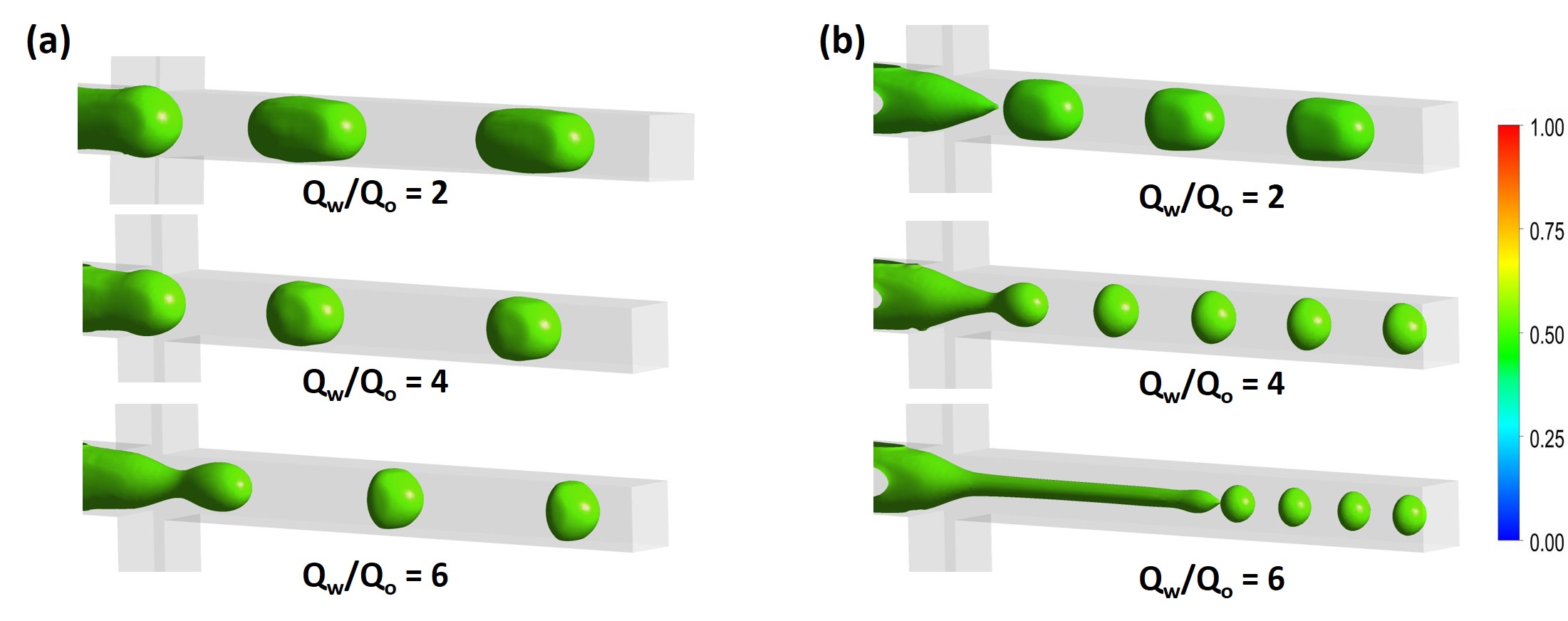}
	\caption{\label{fig:VVA1} Three dimensional iso-surface evolution of droplet shapes for various flow rate ratio in (a) oil\textendash water, and (b) oil\textendash water+40\% glycerol system at $Q_o$= 400~$\mu$L/min.}
\end{figure}
From Fig.~\ref{fig:VVA1}a, it is observed that with increasing continuous phase flow rate, dripping regime is not influenced for oil-water system. However, droplet shape significantly changes from slug to near spherical at a higher flow rate. However, in oil\textendash water+40\% glycerol system, dripping regime is observed at a lower flow rate, as shown in Fig.~\ref{fig:VVA1}b. With increasing flow rate, the regime shifts from dripping to jetting and considerable change in droplet shape is observed, as depicted in Fig.~\ref{fig:VVA1}b. 

\begin{figure}[!ht]
	\centering
	\includegraphics[width=\linewidth]{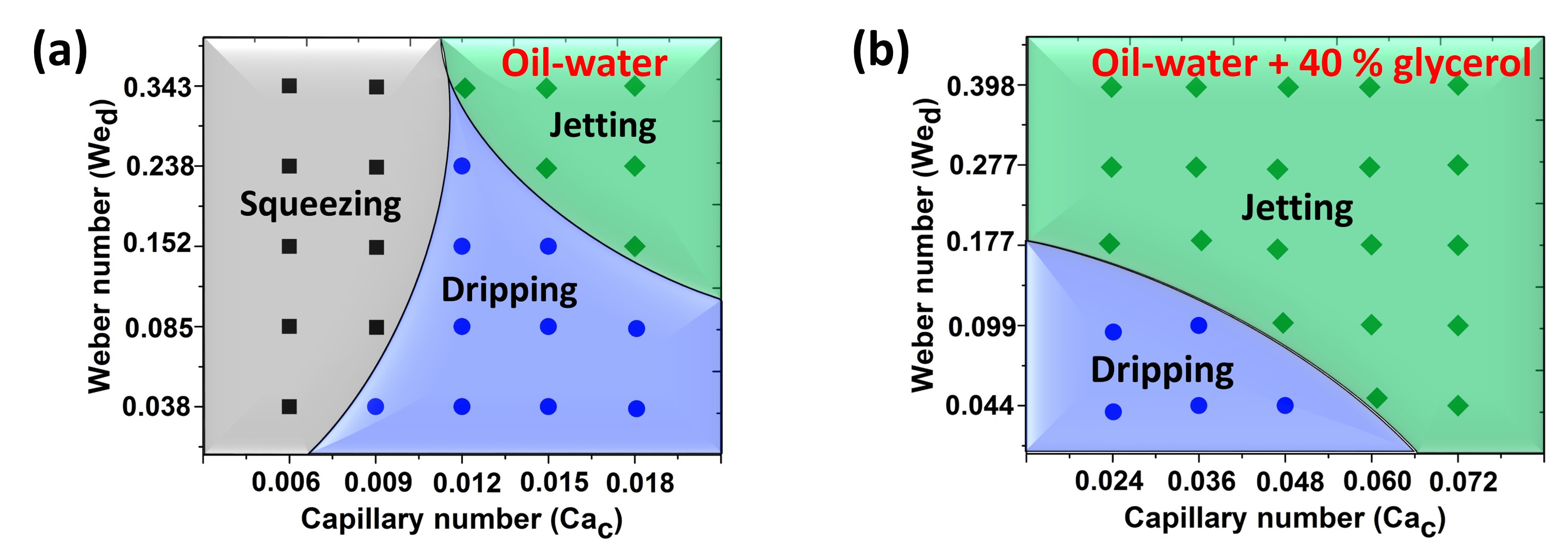}
	\caption{\label{fig:VR1} Flow pattern maps for different liquid-liquid systems in a flow-focusing microchannel containing (a) oil\textendash water, and (b) oil\textendash water+40\% glycerol.}
\end{figure}

\begin{figure}[!ht]
	\centering
	\includegraphics[width=\linewidth]{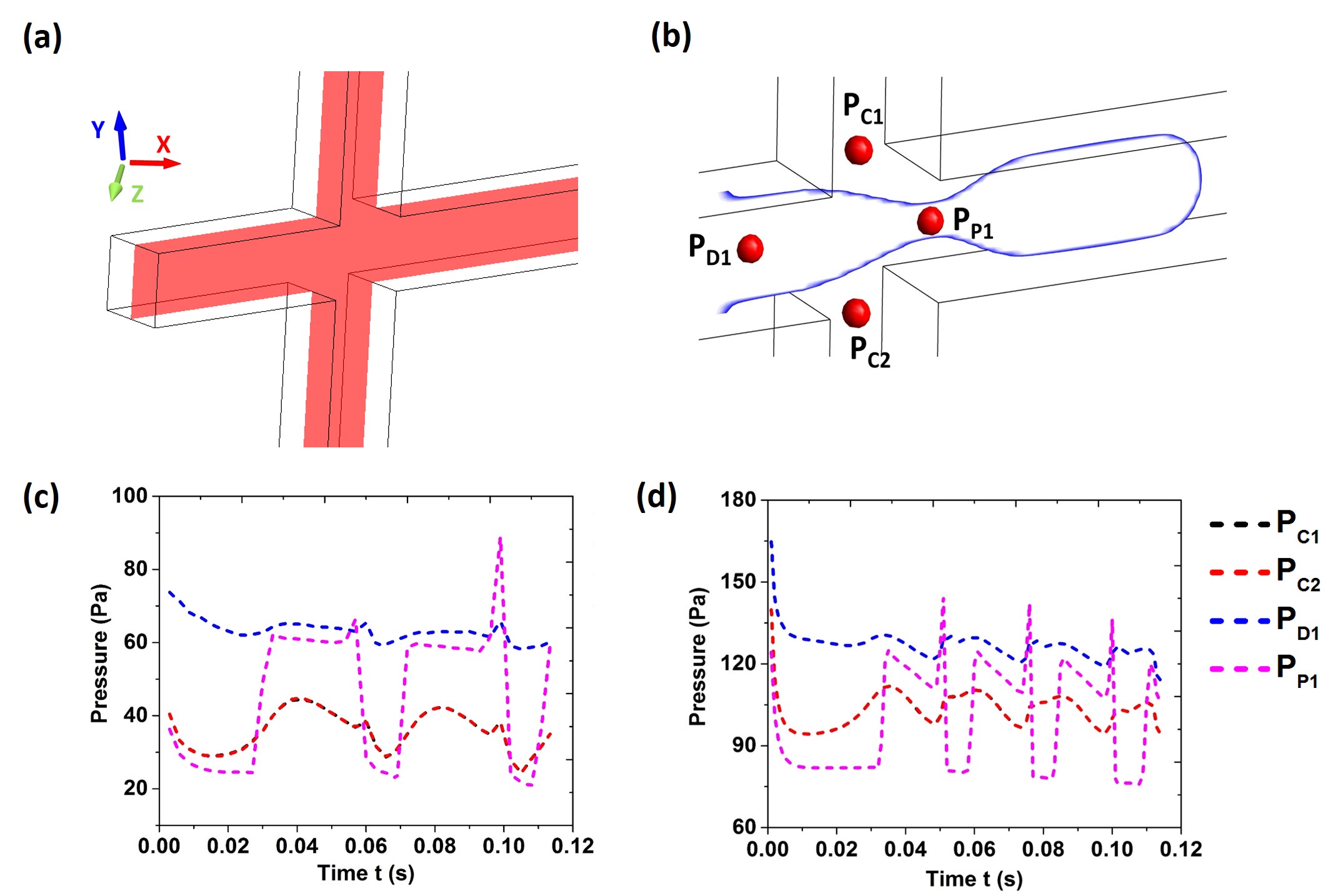}
	\caption{\label{fig:V29} (a) XY plane in the microchannel at $Z=300$ $\mu$m, (b) locations of four selected pressure recording points on the XY plane, (c) evolution of pressures in the microchannel for oil\textendash water, and (d) oil\textendash water+40\% glycerol system at fixed operating conditions of $\theta$= 120\textdegree, $Q_w/Q_o=2$ and $Q_w$= 400~$\mu$L/min.}	
\end{figure}

Furthermore, the flow regime maps are developed for oil-water and oil-water+40\% glycerol systems as a function of continuous phase Capillary number ($Ca_{c}=\eta_{c}$$U_{c}$/$\gamma$) and dispersed phase Weber number ($We_{d}=\rho U_{d}^{2} W/\gamma$) by altering the corresponding phase flow rates. As discussed earlier, three different flow regimes namely squeezing, dripping, and jetting patterns are identified. Fig.~\ref{fig:VR1}a shows that the oil-water system has larger squeezing and dripping regimes than the oil-water+40\% glycerol system, where squeezing regime is completely absent in the range of our studied operating condition.     
In addition, pressure evolution in the continuous and dispersed phases, as well as at the pinch-off position is analyzed during the droplet formation process by creating a XY plane in the middle of the microchannel at $Z=300$ $\mu$m (Fig.~\ref{fig:V29}a).
Fig.~\ref{fig:V29}b shows four points ($ P_{C1}, P_{C2},P_{D1}$, and $P_{P1}$), which are chosen on that XY plane. $P_{C1}$~(2100~$\mu$m, 700~$\mu$m) and $P_{C2}$~(2100~$\mu$m, -100~$\mu$m) are located at the upstream of side channels for the continuous phase. $ P_{D1}$~(1500~$\mu$m, 300~$\mu$m) and $P_{P1}$~(2500~$\mu$m, 300~$\mu$m) are considered in the dispersed phase and at the downstream edge of the cross junction, respectively. As the dispersed phase enters the cross junction and its tip partially blocks the main channel, the upstream and downstream pressure of the continuous phase ($P_{C1}$ and $ P_{C2}$) increases and reaches a maximum value when the dispersed phase is completely occupied at the cross junction. Subsequently, $P_{C1}$ and $ P_{C2}$ decrease during the blocking time until the neck breakup occurs, as shown in Fig.~\ref{fig:V29}b. Afterward, $ P_{C1}$ and $ P_{C2}$ drop suddenly and a sharp peak can be seen for $P_{D1}$. This is attributed to the rebound after separation of the droplet. Once the droplet pinch-off occurs, $ P_{C}$ and $ P_{D}$ regain their pressure for the next cycle. It is worth noting that $ P_{P1}$ indicates dispersed phase pressure at the droplet pinch-off position. Likewise, pressure fluctuations are evidenced for oil\textendash water+40\% glycerol, as depicted in Fig.~\ref{fig:V29}c. The pressure fluctuation in oil\textendash water+40\% glycerol is higher in magnitude as compared to the oil-water system due to higher viscous resistance.

\section{Conclusions}
A three-dimensional computational study on droplet formation and breakup dynamics in viscous liquids is carried out in a flow-focusing device using CLSVOF method. Oil is considered as dispersed phase and water/glycerol solutions are taken as continuous phase. The effects of viscosity ratio, interfacial tension, and flow rate ratio are investigated in detail. For all cases, the liquid film thickness around the droplet is precisely captured to understand its effect on the droplet velocity. Results show that droplet length and volume decrease with increasing viscosity and flow rate ratios; however, droplet velocity and liquid film thickness increase. Scaling laws are proposed to determine the droplet length based on $Ca$. Dripping regime is observed at lower $Ca$ and flow rate ratios. However, jetting regime is found at the higher flow rate ratios. Flow regime transition from dripping to jetting is observed with changing interfacial tension, flow rate ratio, and dispersed phase viscosity. The impact of continuous phase velocity on dispersed phase shearing and droplet breakup is demonstrated. Velocity field variations inside the droplet and liquid slug are also analyzed for various systems. A parameter defined as droplet deformation index is used to identify the shape of droplets, which shows that near spherical droplets are typically formed in all cases of jetting regime. Plug-shaped droplets are obtained in the dripping regime. The evolution of pressure over time in the continuous and dispersed phases is analyzed. From the model predictions, a unified scaling law is proposed to estimate the droplet length for a wide range (0.006-0.072) of $Ca$. Flow regime maps for two different liquid-liquid systems are also developed as a function of continuous phase Capillary number and dispersed phase Weber number. Nonetheless, it is of considerable importance to identify the liquid film thickness around the droplet, which essentially dictates the droplet dynamics. None of the previous studies had clearly delineated this influence, whereas our results herein are demonstrated based on this fundamental aspect. Therefore, this work contributes toward accurate predictions of droplet formation mechanism in various flow regimes. These outcomes are likely to guide experiments, where monosized droplet generation and its manipulation with tunable volume are essential.    


\section*{Acknowledgments}
This work is supported by IIT Kharagpur under the Centre of Excellence for Training and Research in Microfluidics.

\section*{Nomenclature}
\noindent $Ca$ = Capillary number (=~$\eta U/\gamma$)\\
$We$ = Weber number (=~$\rho U_{d}^{2} W/\gamma$)\\
$W$ = width ($m$) \\
$Q$ = flow rate ($m^{3}/s$)\\ 
$U$  =  velocity (m/s) \\
$V$ = volume ($m^{3}$)\\
$\hat{N}$	= unit normal vector\\
$p$ = pressure (Pa)\\
$t$ = flow time (s)\\
$H$ = height (m)\\
$h$ = depth (m)\\
$\textit{H}(\varphi)$ = Heaviside function \\
$\vec{\chi}$ = position vector \\
$\textit{a}$ =  interface thickness (m)\\
$\textit{d}$ =  shortest distance \\
$\delta ( \varphi )$ =  Direct delta function \\ 
$D.I$ = droplet defromation index \\
\textit{Greek symbols}\\	
$\alpha$  = volume fraction\\
$\dot{\gamma } $ = shear rate (1/s)\\
$\delta$ = liquid film thickness (m) \\ 
$\theta$ = contact angle (\textdegree)\\
$\kappa_{n}$ = radius of curvature \\
$\eta$ = dynamic viscosity (kg/m.s)\\
$\rho$  = density ($kg/m^{3}$)\\
$\gamma$  = interfacial tension (N/m)\\ 
$\overline{\overline\tau}$ = shear stress (Pa)\\
$\varphi $  = level set function \\
\textit{Subscripts}\\
$D$ = droplet \\
$c$ = continuous phase\\
$d$ = dispersed phase\\
$o$ = oil\\
$w$ = water


\begin{mcitethebibliography}{63}
	\providecommand*\natexlab[1]{#1}
	\providecommand*\mciteSetBstSublistMode[1]{}
	\providecommand*\mciteSetBstMaxWidthForm[2]{}
	\providecommand*\mciteBstWouldAddEndPuncttrue
	{\def\EndOfBibitem{\unskip.}}
	\providecommand*\mciteBstWouldAddEndPunctfalse
	{\let\EndOfBibitem\relax}
	\providecommand*\mciteSetBstMidEndSepPunct[3]{}
	\providecommand*\mciteSetBstSublistLabelBeginEnd[3]{}
	\providecommand*\EndOfBibitem{}
	\mciteSetBstSublistMode{f}
	\mciteSetBstMaxWidthForm{subitem}{(\alph{mcitesubitemcount})}
	\mciteSetBstSublistLabelBeginEnd
	{\mcitemaxwidthsubitemform\space}
	{\relax}
	{\relax}
	
	\bibitem[Yu \latin{et~al.}(2015)Yu, Cheng, Zhou, and Zheng]{yu2015demand}
	Yu,~X.; Cheng,~G.; Zhou,~M.-D.; Zheng,~S.-Y. On-demand one-step synthesis of
	monodisperse functional polymeric microspheres with droplet microfluidics.
	\emph{Langmuir} \textbf{2015}, \emph{31}, 3982\relax
	\mciteBstWouldAddEndPuncttrue
	\mciteSetBstMidEndSepPunct{\mcitedefaultmidpunct}
	{\mcitedefaultendpunct}{\mcitedefaultseppunct}\relax
	\EndOfBibitem
	\bibitem[Dang \latin{et~al.}(2013)Dang, Cheney, Qian, Joo, and
	Min]{dang2013reactivity}
	Dang,~T.-D.; Cheney,~M.~A.; Qian,~S.; Joo,~S.~W.; Min,~B.-K. Reactivity of poly
	(dimethylsiloxane) toward acidic permanganate. \emph{J. Ind. Eng. Chem.}
	\textbf{2013}, \emph{19}, 1770\relax
	\mciteBstWouldAddEndPuncttrue
	\mciteSetBstMidEndSepPunct{\mcitedefaultmidpunct}
	{\mcitedefaultendpunct}{\mcitedefaultseppunct}\relax
	\EndOfBibitem
	\bibitem[Bennett \latin{et~al.}(2018)Bennett, Kristof, Vasudevan, Genzer,
	Srogl, and Abolhasani]{bennett2018microfluidic}
	Bennett,~J.~A.; Kristof,~A.~J.; Vasudevan,~V.; Genzer,~J.; Srogl,~J.;
	Abolhasani,~M. Microfluidic synthesis of elastomeric microparticles: A case
	study in catalysis of palladium-mediated cross-coupling. \emph{AIChE J.}
	\textbf{2018}, \emph{64}, 3188\relax
	\mciteBstWouldAddEndPuncttrue
	\mciteSetBstMidEndSepPunct{\mcitedefaultmidpunct}
	{\mcitedefaultendpunct}{\mcitedefaultseppunct}\relax
	\EndOfBibitem
	\bibitem[Nieves-Remacha \latin{et~al.}(2012)Nieves-Remacha, Kulkarni, and
	Jensen]{nieves2012hydrodynamics}
	Nieves-Remacha,~M.~J.; Kulkarni,~A.~A.; Jensen,~K.~F. Hydrodynamics of
	liquid--liquid dispersion in an advanced-flow reactor. \emph{Ind. Eng. Chem.
		Res.} \textbf{2012}, \emph{51}, 16251\relax
	\mciteBstWouldAddEndPuncttrue
	\mciteSetBstMidEndSepPunct{\mcitedefaultmidpunct}
	{\mcitedefaultendpunct}{\mcitedefaultseppunct}\relax
	\EndOfBibitem
	\bibitem[Song \latin{et~al.}(2006)Song, Chen, and Ismagilov]{song2006}
	Song,~H.; Chen,~D.~L.; Ismagilov,~R.~F. Reactions in droplets in microfluidic
	channels. \emph{Angew. Chem. Int. Ed.} \textbf{2006}, \emph{45}, 7336\relax
	\mciteBstWouldAddEndPuncttrue
	\mciteSetBstMidEndSepPunct{\mcitedefaultmidpunct}
	{\mcitedefaultendpunct}{\mcitedefaultseppunct}\relax
	\EndOfBibitem
	\bibitem[Haeberle and Zengerle(2007)Haeberle, and Zengerle]{haeberle2007}
	Haeberle,~S.; Zengerle,~R. Microfluidic platforms for lab-on-a-chip
	applications. \emph{Lab. Chip} \textbf{2007}, \emph{7}, 1094\relax
	\mciteBstWouldAddEndPuncttrue
	\mciteSetBstMidEndSepPunct{\mcitedefaultmidpunct}
	{\mcitedefaultendpunct}{\mcitedefaultseppunct}\relax
	\EndOfBibitem
	\bibitem[Chueluecha \latin{et~al.}(2017)Chueluecha, Kaewchada, and
	Jaree]{chueluecha2017enhancement}
	Chueluecha,~N.; Kaewchada,~A.; Jaree,~A. Enhancement of biodiesel synthesis
	using co-solvent in a packed-microchannel. \emph{J. Ind. Eng. Chem.}
	\textbf{2017}, \emph{51}, 162\relax
	\mciteBstWouldAddEndPuncttrue
	\mciteSetBstMidEndSepPunct{\mcitedefaultmidpunct}
	{\mcitedefaultendpunct}{\mcitedefaultseppunct}\relax
	\EndOfBibitem
	\bibitem[Campbell \latin{et~al.}(2018)Campbell, Parker, Bennett, Yusuf,
	Al-Rashdi, Lustik, Li, and Abolhasani]{campbell2018continuous}
	Campbell,~Z.~S.; Parker,~M.; Bennett,~J.~A.; Yusuf,~S.; Al-Rashdi,~A.~K.;
	Lustik,~J.; Li,~F.; Abolhasani,~M. Continuous Synthesis of Monodisperse
	Yolk-Shell Titania Microspheres. \emph{Chem. Mater.} \textbf{2018},
	\emph{30}, 8948\relax
	\mciteBstWouldAddEndPuncttrue
	\mciteSetBstMidEndSepPunct{\mcitedefaultmidpunct}
	{\mcitedefaultendpunct}{\mcitedefaultseppunct}\relax
	\EndOfBibitem
	\bibitem[Chand \latin{et~al.}(2015)Chand, Jeun, Park, Kim, Shin, and
	Kim]{chand2015electroimmobilization}
	Chand,~R.; Jeun,~J.-H.; Park,~M.-H.; Kim,~J.-M.; Shin,~I.-S.; Kim,~Y.-S.
	Electroimmobilization of DNA for ultrafast detection on a microchannel
	integrated pentacene TFT. \emph{J. Ind. Eng. Chem.} \textbf{2015}, \emph{21},
	126\relax
	\mciteBstWouldAddEndPuncttrue
	\mciteSetBstMidEndSepPunct{\mcitedefaultmidpunct}
	{\mcitedefaultendpunct}{\mcitedefaultseppunct}\relax
	\EndOfBibitem
	\bibitem[Abolhasani and Jensen(2016)Abolhasani, and Jensen]{abolhasani2016}
	Abolhasani,~M.; Jensen,~K.~F. Oscillatory multiphase flow strategy for
	chemistry and biology. \emph{Lab. Chip} \textbf{2016}, \emph{16}, 2775\relax
	\mciteBstWouldAddEndPuncttrue
	\mciteSetBstMidEndSepPunct{\mcitedefaultmidpunct}
	{\mcitedefaultendpunct}{\mcitedefaultseppunct}\relax
	\EndOfBibitem
	\bibitem[Ma \latin{et~al.}(2018)Ma, Zhang, Fu, Zhu, Wang, Ma, and
	Luo]{ma2018manipulation}
	Ma,~R.; Zhang,~Q.; Fu,~T.; Zhu,~C.; Wang,~K.; Ma,~Y.; Luo,~G. Manipulation of
	microdroplets at a T-junction: Coalescence and scaling law. \emph{J. Ind.
		Eng. Chem.} \textbf{2018}, \emph{65}, 272\relax
	\mciteBstWouldAddEndPuncttrue
	\mciteSetBstMidEndSepPunct{\mcitedefaultmidpunct}
	{\mcitedefaultendpunct}{\mcitedefaultseppunct}\relax
	\EndOfBibitem
	\bibitem[Teh \latin{et~al.}(2008)Teh, Lin, Hung, and Lee]{teh2008droplet}
	Teh,~S.-Y.; Lin,~R.; Hung,~L.-H.; Lee,~A.~P. Droplet microfluidics. \emph{Lab.
		Chip} \textbf{2008}, \emph{8}, 198\relax
	\mciteBstWouldAddEndPuncttrue
	\mciteSetBstMidEndSepPunct{\mcitedefaultmidpunct}
	{\mcitedefaultendpunct}{\mcitedefaultseppunct}\relax
	\EndOfBibitem
	\bibitem[Bordbar \latin{et~al.}(2018)Bordbar, Taassob, Zarnaghsh, and
	Kamali]{bordbar2018slug}
	Bordbar,~A.; Taassob,~A.; Zarnaghsh,~A.; Kamali,~R. Slug flow in microchannels:
	Numerical simulation and applications. \emph{J. Ind. Eng. Chem.}
	\textbf{2018}, \emph{62}, 26\relax
	\mciteBstWouldAddEndPuncttrue
	\mciteSetBstMidEndSepPunct{\mcitedefaultmidpunct}
	{\mcitedefaultendpunct}{\mcitedefaultseppunct}\relax
	\EndOfBibitem
	\bibitem[Anna \latin{et~al.}(2003)Anna, Bontoux, and Stone]{anna2003formation}
	Anna,~S.~L.; Bontoux,~N.; Stone,~H.~A. Formation of dispersions using flow
	focusing in microchannels. \emph{Appl. Phys. Lett.} \textbf{2003}, \emph{82},
	364\relax
	\mciteBstWouldAddEndPuncttrue
	\mciteSetBstMidEndSepPunct{\mcitedefaultmidpunct}
	{\mcitedefaultendpunct}{\mcitedefaultseppunct}\relax
	\EndOfBibitem
	\bibitem[Dang \latin{et~al.}(2012)Dang, Kim, Kim, and Kim]{dang2012preparation}
	Dang,~T.-D.; Kim,~Y.~H.; Kim,~H.~G.; Kim,~G.~M. Preparation of monodisperse PEG
	hydrogel microparticles using a microfluidic flow-focusing device. \emph{J.
		Ind. Eng. Chem.} \textbf{2012}, \emph{18}, 1308\relax
	\mciteBstWouldAddEndPuncttrue
	\mciteSetBstMidEndSepPunct{\mcitedefaultmidpunct}
	{\mcitedefaultendpunct}{\mcitedefaultseppunct}\relax
	\EndOfBibitem
	\bibitem[Tan \latin{et~al.}(2008)Tan, Xu, Li, and Luo]{tan2008drop}
	Tan,~J.; Xu,~J.; Li,~S.; Luo,~G. Drop dispenser in a cross-junction
	microfluidic device: Scaling and mechanism of break-up. \emph{Chem. Eng. J.}
	\textbf{2008}, \emph{136}, 306\relax
	\mciteBstWouldAddEndPuncttrue
	\mciteSetBstMidEndSepPunct{\mcitedefaultmidpunct}
	{\mcitedefaultendpunct}{\mcitedefaultseppunct}\relax
	\EndOfBibitem
	\bibitem[Fu \latin{et~al.}(2012)Fu, Wu, Ma, and Li]{fu-2012s}
	Fu,~T.; Wu,~Y.; Ma,~Y.; Li,~H.~Z. Droplet formation and breakup dynamics in
	microfluidic flow-focusing devices: from dripping to jetting. \emph{Chem.
		Eng. Sci.} \textbf{2012}, \emph{84}, 207\relax
	\mciteBstWouldAddEndPuncttrue
	\mciteSetBstMidEndSepPunct{\mcitedefaultmidpunct}
	{\mcitedefaultendpunct}{\mcitedefaultseppunct}\relax
	\EndOfBibitem
	\bibitem[Ma \latin{et~al.}(2014)Ma, Sherwood, Huck, and Balabani]{ma2014flow}
	Ma,~S.; Sherwood,~J.~M.; Huck,~W.~T.; Balabani,~S. On the flow topology inside
	droplets moving in rectangular microchannels. \emph{Lab. Chip} \textbf{2014},
	\emph{14}, 3611\relax
	\mciteBstWouldAddEndPuncttrue
	\mciteSetBstMidEndSepPunct{\mcitedefaultmidpunct}
	{\mcitedefaultendpunct}{\mcitedefaultseppunct}\relax
	\EndOfBibitem
	\bibitem[Chen \latin{et~al.}(2015)Chen, Glawdel, Cui, and Ren]{chen2015model}
	Chen,~X.; Glawdel,~T.; Cui,~N.; Ren,~C.~L. Model of droplet generation in flow
	focusing generators operating in the squeezing regime. \emph{Microfluid.
		Nanofluid} \textbf{2015}, \emph{18}, 1341\relax
	\mciteBstWouldAddEndPuncttrue
	\mciteSetBstMidEndSepPunct{\mcitedefaultmidpunct}
	{\mcitedefaultendpunct}{\mcitedefaultseppunct}\relax
	\EndOfBibitem
	\bibitem[Kovalchuk \latin{et~al.}(2018)Kovalchuk, Roumpea, Nowak, Chinaud,
	Angeli, and Simmons]{kovalchuk2018effect}
	Kovalchuk,~N.~M.; Roumpea,~E.; Nowak,~E.; Chinaud,~M.; Angeli,~P.;
	Simmons,~M.~J. Effect of surfactant on emulsification in microchannels.
	\emph{Chem. Eng. Sci.} \textbf{2018}, \emph{176}, 139\relax
	\mciteBstWouldAddEndPuncttrue
	\mciteSetBstMidEndSepPunct{\mcitedefaultmidpunct}
	{\mcitedefaultendpunct}{\mcitedefaultseppunct}\relax
	\EndOfBibitem
	\bibitem[Wu \latin{et~al.}(2017)Wu, Cao, and Sund{\'e}n]{wu2017liquid}
	Wu,~Z.; Cao,~Z.; Sund{\'e}n,~B. Liquid-liquid flow patterns and slug
	hydrodynamics in square microchannels of cross-shaped junctions. \emph{Chem.
		Eng. Sci.} \textbf{2017}, \emph{174}, 56\relax
	\mciteBstWouldAddEndPuncttrue
	\mciteSetBstMidEndSepPunct{\mcitedefaultmidpunct}
	{\mcitedefaultendpunct}{\mcitedefaultseppunct}\relax
	\EndOfBibitem
	\bibitem[Wu \latin{et~al.}(2013)Wu, Fu, Ma, and Li]{wu2013ferrofluid}
	Wu,~Y.; Fu,~T.; Ma,~Y.; Li,~H.~Z. Ferrofluid droplet formation and breakup
	dynamics in a microfluidic flow-focusing device. \emph{Soft Matter}
	\textbf{2013}, \emph{9}, 9792\relax
	\mciteBstWouldAddEndPuncttrue
	\mciteSetBstMidEndSepPunct{\mcitedefaultmidpunct}
	{\mcitedefaultendpunct}{\mcitedefaultseppunct}\relax
	\EndOfBibitem
	\bibitem[Varma \latin{et~al.}(2017)Varma, Wu, Wang, and
	Ramanujan]{varma2017magnetic}
	Varma,~V.; Wu,~R.; Wang,~Z.; Ramanujan,~R. Magnetic Janus particles synthesized
	using droplet micro-magnetofluidic techniques for protein detection.
	\emph{Lab. Chip} \textbf{2017}, \emph{17}, 3514\relax
	\mciteBstWouldAddEndPuncttrue
	\mciteSetBstMidEndSepPunct{\mcitedefaultmidpunct}
	{\mcitedefaultendpunct}{\mcitedefaultseppunct}\relax
	\EndOfBibitem
	\bibitem[Du \latin{et~al.}(2016)Du, Fu, Zhang, Zhu, Ma, and Li]{du2016}
	Du,~W.; Fu,~T.; Zhang,~Q.; Zhu,~C.; Ma,~Y.; Li,~H.~Z. Breakup dynamics for
	droplet formation in a flow-focusing device: Rupture position of viscoelastic
	thread from matrix. \emph{Chem. Eng. Sci.} \textbf{2016}, \emph{153},
	255\relax
	\mciteBstWouldAddEndPuncttrue
	\mciteSetBstMidEndSepPunct{\mcitedefaultmidpunct}
	{\mcitedefaultendpunct}{\mcitedefaultseppunct}\relax
	\EndOfBibitem
	\bibitem[Sonthalia \latin{et~al.}(2016)Sonthalia, Ng, and
	Ramachandran]{sonthalia2016formation}
	Sonthalia,~R.; Ng,~S.; Ramachandran,~A. Formation of extremely fine water
	droplets in sheared, concentrated bitumen solutions via surfactant-mediated
	tip streaming. \emph{Fuel} \textbf{2016}, \emph{180}, 538\relax
	\mciteBstWouldAddEndPuncttrue
	\mciteSetBstMidEndSepPunct{\mcitedefaultmidpunct}
	{\mcitedefaultendpunct}{\mcitedefaultseppunct}\relax
	\EndOfBibitem
	\bibitem[Zhou \latin{et~al.}(2006)Zhou, Yue, and Feng]{zhou2006formation}
	Zhou,~C.; Yue,~P.; Feng,~J.~J. Formation of simple and compound drops in
	microfluidic devices. \emph{Phys. Fluids} \textbf{2006}, \emph{18},
	092105\relax
	\mciteBstWouldAddEndPuncttrue
	\mciteSetBstMidEndSepPunct{\mcitedefaultmidpunct}
	{\mcitedefaultendpunct}{\mcitedefaultseppunct}\relax
	\EndOfBibitem
	\bibitem[Bai \latin{et~al.}(2017)Bai, He, Yang, Zhou, and Wang]{bai2017three}
	Bai,~F.; He,~X.; Yang,~X.; Zhou,~R.; Wang,~C. Three dimensional phase-field
	investigation of droplet formation in microfluidic flow focusing devices with
	experimental validation. \emph{Int. J. Multiphase Flow} \textbf{2017},
	\emph{93}, 130\relax
	\mciteBstWouldAddEndPuncttrue
	\mciteSetBstMidEndSepPunct{\mcitedefaultmidpunct}
	{\mcitedefaultendpunct}{\mcitedefaultseppunct}\relax
	\EndOfBibitem
	\bibitem[Wu \latin{et~al.}(2008)Wu, Tsutahara, Kim, and Ha]{wu2008three}
	Wu,~L.; Tsutahara,~M.; Kim,~L.~S.; Ha,~M. Three-dimensional lattice Boltzmann
	simulations of droplet formation in a cross-junction microchannel. \emph{Int.
		J. Multiphase Flow} \textbf{2008}, \emph{34}, 852\relax
	\mciteBstWouldAddEndPuncttrue
	\mciteSetBstMidEndSepPunct{\mcitedefaultmidpunct}
	{\mcitedefaultendpunct}{\mcitedefaultseppunct}\relax
	\EndOfBibitem
	\bibitem[Gupta \latin{et~al.}(2014)Gupta, Matharoo, Makkar, and
	Kumar]{gupta2014droplet}
	Gupta,~A.; Matharoo,~H.~S.; Makkar,~D.; Kumar,~R. Droplet formation via
	squeezing mechanism in a microfluidic flow-focusing device. \emph{Comput.
		Fluids} \textbf{2014}, \emph{100}, 218\relax
	\mciteBstWouldAddEndPuncttrue
	\mciteSetBstMidEndSepPunct{\mcitedefaultmidpunct}
	{\mcitedefaultendpunct}{\mcitedefaultseppunct}\relax
	\EndOfBibitem
	\bibitem[Ong \latin{et~al.}(2007)Ong, Hua, Zhang, Teo, Zhuo, Nguyen,
	Ranganathan, and Yobas]{ong2007experimental}
	Ong,~W.-L.; Hua,~J.; Zhang,~B.; Teo,~T.-Y.; Zhuo,~J.; Nguyen,~N.-T.;
	Ranganathan,~N.; Yobas,~L. Experimental and computational analysis of droplet
	formation in a high-performance flow-focusing geometry. \emph{Sens. Actuators
		A Phys.} \textbf{2007}, \emph{138}, 203\relax
	\mciteBstWouldAddEndPuncttrue
	\mciteSetBstMidEndSepPunct{\mcitedefaultmidpunct}
	{\mcitedefaultendpunct}{\mcitedefaultseppunct}\relax
	\EndOfBibitem
	\bibitem[Hoang \latin{et~al.}(2013)Hoang, van Steijn, Portela, Kreutzer, and
	Kleijn]{hoang2013benchmark}
	Hoang,~D.~A.; van Steijn,~V.; Portela,~L.~M.; Kreutzer,~M.~T.; Kleijn,~C.~R.
	Benchmark numerical simulations of segmented two-phase flows in microchannels
	using the Volume of Fluid method. \emph{Comput. Fluids} \textbf{2013},
	\emph{86}, 28\relax
	\mciteBstWouldAddEndPuncttrue
	\mciteSetBstMidEndSepPunct{\mcitedefaultmidpunct}
	{\mcitedefaultendpunct}{\mcitedefaultseppunct}\relax
	\EndOfBibitem
	\bibitem[Li \latin{et~al.}(2015)Li, Jain, Ma, and Nandakumar]{li2015control}
	Li,~Y.; Jain,~M.; Ma,~Y.; Nandakumar,~K. Control of the breakup process of
	viscous droplets by an external electric field inside a microfluidic device.
	\emph{Soft Matter} \textbf{2015}, \emph{11}, 3884\relax
	\mciteBstWouldAddEndPuncttrue
	\mciteSetBstMidEndSepPunct{\mcitedefaultmidpunct}
	{\mcitedefaultendpunct}{\mcitedefaultseppunct}\relax
	\EndOfBibitem
	\bibitem[Lan \latin{et~al.}(2014)Lan, Li, Wang, and Luo]{lan2014cfd}
	Lan,~W.; Li,~S.; Wang,~Y.; Luo,~G. CFD simulation of droplet formation in
	microchannels by a modified level set method. \emph{Ind. Eng. Chem. Res.}
	\textbf{2014}, \emph{53}, 4913\relax
	\mciteBstWouldAddEndPuncttrue
	\mciteSetBstMidEndSepPunct{\mcitedefaultmidpunct}
	{\mcitedefaultendpunct}{\mcitedefaultseppunct}\relax
	\EndOfBibitem
	\bibitem[Popinet and Zaleski(1999)Popinet, and Zaleski]{popinet1999}
	Popinet,~S.; Zaleski,~S. A front-tracking algorithm for accurate representation
	of surface tension. \emph{Int. J. Numer. Meth. Fl.} \textbf{1999}, \emph{30},
	775\relax
	\mciteBstWouldAddEndPuncttrue
	\mciteSetBstMidEndSepPunct{\mcitedefaultmidpunct}
	{\mcitedefaultendpunct}{\mcitedefaultseppunct}\relax
	\EndOfBibitem
	\bibitem[Harvie \latin{et~al.}(2006)Harvie, Davidson, and Rudman]{harvie2006}
	Harvie,~D.~J.; Davidson,~M.; Rudman,~M. An analysis of parasitic current
	generation in volume of fluid simulations. \emph{Appl. Math. Model.}
	\textbf{2006}, \emph{30}, 1056\relax
	\mciteBstWouldAddEndPuncttrue
	\mciteSetBstMidEndSepPunct{\mcitedefaultmidpunct}
	{\mcitedefaultendpunct}{\mcitedefaultseppunct}\relax
	\EndOfBibitem
	\bibitem[Popinet(2003)]{popinet2003gerris}
	Popinet,~S. Gerris: A tree-based adaptive solver for the incompressible Euler
	equations in complex geometries. \emph{J. Comput. Phys.} \textbf{2003},
	\emph{190}, 572\relax
	\mciteBstWouldAddEndPuncttrue
	\mciteSetBstMidEndSepPunct{\mcitedefaultmidpunct}
	{\mcitedefaultendpunct}{\mcitedefaultseppunct}\relax
	\EndOfBibitem
	\bibitem[Francois \latin{et~al.}(2006)Francois, Cummins, Dendy, Kothe,
	Sicilian, and Williams]{francois2006balanced}
	Francois,~M.~M.; Cummins,~S.~J.; Dendy,~E.~D.; Kothe,~D.~B.; Sicilian,~J.~M.;
	Williams,~M.~W. A balanced-force algorithm for continuous and sharp
	interfacial surface tension models within a volume tracking framework.
	\emph{J. Comput. Phys.} \textbf{2006}, \emph{213}, 141\relax
	\mciteBstWouldAddEndPuncttrue
	\mciteSetBstMidEndSepPunct{\mcitedefaultmidpunct}
	{\mcitedefaultendpunct}{\mcitedefaultseppunct}\relax
	\EndOfBibitem
	\bibitem[W{\"o}rner(2012)]{Worner2012}
	W{\"o}rner,~M. Numerical modeling of multiphase flows in microfluidics and
	micro process engineering: a review of methods and applications.
	\emph{Microfluid. Nanofluid.} \textbf{2012}, \emph{12}, 841\relax
	\mciteBstWouldAddEndPuncttrue
	\mciteSetBstMidEndSepPunct{\mcitedefaultmidpunct}
	{\mcitedefaultendpunct}{\mcitedefaultseppunct}\relax
	\EndOfBibitem
	\bibitem[Denner and van Wachem(2014)Denner, and van Wachem]{denner2014fully}
	Denner,~F.; van Wachem,~B.~G. Fully-coupled balanced-force VOF framework for
	arbitrary meshes with least-squares curvature evaluation from volume
	fractions. \emph{Numer. Heat Transfer B Fund.} \textbf{2014}, \emph{65},
	218\relax
	\mciteBstWouldAddEndPuncttrue
	\mciteSetBstMidEndSepPunct{\mcitedefaultmidpunct}
	{\mcitedefaultendpunct}{\mcitedefaultseppunct}\relax
	\EndOfBibitem
	\bibitem[Guo \latin{et~al.}(2015)Guo, Fletcher, and Haynes]{guo2015}
	Guo,~Z.; Fletcher,~D.~F.; Haynes,~B.~S. Implementation of a height function
	method to alleviate spurious currents in CFD modelling of annular flow in
	microchannels. \emph{Appl. Math. Model.} \textbf{2015}, \emph{39}, 4665\relax
	\mciteBstWouldAddEndPuncttrue
	\mciteSetBstMidEndSepPunct{\mcitedefaultmidpunct}
	{\mcitedefaultendpunct}{\mcitedefaultseppunct}\relax
	\EndOfBibitem
	\bibitem[Sussman and Puckett(2000)Sussman, and Puckett]{sussman2000}
	Sussman,~M.; Puckett,~E.~G. A coupled level set and volume-of-fluid method for
	computing 3D and axisymmetric incompressible two-phase flows. \emph{J.
		Comput. Phys.} \textbf{2000}, \emph{162}, 301\relax
	\mciteBstWouldAddEndPuncttrue
	\mciteSetBstMidEndSepPunct{\mcitedefaultmidpunct}
	{\mcitedefaultendpunct}{\mcitedefaultseppunct}\relax
	\EndOfBibitem
	\bibitem[Keshavarzi \latin{et~al.}(2014)Keshavarzi, Pawell, Barber, and
	Yeoh]{keshavarzi2014}
	Keshavarzi,~G.; Pawell,~R.~S.; Barber,~T.~J.; Yeoh,~G.~H. Transient analysis of
	a single rising bubble used for numerical validation for multiphase flow.
	\emph{Chem. Eng. Sci.} \textbf{2014}, \emph{112}, 25\relax
	\mciteBstWouldAddEndPuncttrue
	\mciteSetBstMidEndSepPunct{\mcitedefaultmidpunct}
	{\mcitedefaultendpunct}{\mcitedefaultseppunct}\relax
	\EndOfBibitem
	\bibitem[Buwa \latin{et~al.}(2007)Buwa, Gerlach, Durst, and
	Schl{\"u}cker]{buwa2007}
	Buwa,~V.~V.; Gerlach,~D.; Durst,~F.; Schl{\"u}cker,~E. Numerical simulations of
	bubble formation on submerged orifices: period-1 and period-2 bubbling
	regimes. \emph{Chem. Eng. Sci.} \textbf{2007}, \emph{62}, 7119\relax
	\mciteBstWouldAddEndPuncttrue
	\mciteSetBstMidEndSepPunct{\mcitedefaultmidpunct}
	{\mcitedefaultendpunct}{\mcitedefaultseppunct}\relax
	\EndOfBibitem
	\bibitem[Ray \latin{et~al.}(2015)Ray, Biswas, and Sharma]{ray2015}
	Ray,~B.; Biswas,~G.; Sharma,~A. Regimes during liquid drop impact on a liquid
	pool. \emph{J. Fluid Mech.} \textbf{2015}, \emph{768}, 492\relax
	\mciteBstWouldAddEndPuncttrue
	\mciteSetBstMidEndSepPunct{\mcitedefaultmidpunct}
	{\mcitedefaultendpunct}{\mcitedefaultseppunct}\relax
	\EndOfBibitem
	\bibitem[Sontti and Atta(2017)Sontti, and Atta]{sontti2017numerical}
	Sontti,~S.~G.; Atta,~A. Numerical investigation of viscous effect on Taylor
	bubble formation in co-flow microchannel. Computer Aided Chemical
	Engineering. 2017; pp 1201--1206\relax
	\mciteBstWouldAddEndPuncttrue
	\mciteSetBstMidEndSepPunct{\mcitedefaultmidpunct}
	{\mcitedefaultendpunct}{\mcitedefaultseppunct}\relax
	\EndOfBibitem
	\bibitem[Sontti \latin{et~al.}(2019)Sontti, Pallewar, Ghosh, and
	Atta]{sonttiunderstanding}
	Sontti,~S.~G.; Pallewar,~P.~G.; Ghosh,~A.~B.; Atta,~A. Understanding the
	inuence of rheological properties of shear-thinning liquids on segmented ow
	in microchannel using CLSVOF based CFD model. \emph{Can. J. Chem.}
	\textbf{2019}, \emph{97}, 1208\relax
	\mciteBstWouldAddEndPuncttrue
	\mciteSetBstMidEndSepPunct{\mcitedefaultmidpunct}
	{\mcitedefaultendpunct}{\mcitedefaultseppunct}\relax
	\EndOfBibitem
	\bibitem[Sontti and Atta(2018)Sontti, and Atta]{sontti2018formation}
	Sontti,~S.~G.; Atta,~A. Formation characteristics of Taylor bubbles in
	power-law liquids flowing through a microfluidic co-flow device. \emph{J.
		Ind. Eng. Chem.} \textbf{2018}, \emph{65}, 82\relax
	\mciteBstWouldAddEndPuncttrue
	\mciteSetBstMidEndSepPunct{\mcitedefaultmidpunct}
	{\mcitedefaultendpunct}{\mcitedefaultseppunct}\relax
	\EndOfBibitem
	\bibitem[Chakraborty \latin{et~al.}(2016)Chakraborty, Rubio-Rubio, Sevilla, and
	Gordillo]{chakraborty2016}
	Chakraborty,~I.; Rubio-Rubio,~M.; Sevilla,~A.; Gordillo,~J. Numerical
	simulation of axisymmetric drop formation using a coupled level set and
	volume of fluid method. \emph{Int. J. Multiphase Flow} \textbf{2016},
	\emph{84}, 54\relax
	\mciteBstWouldAddEndPuncttrue
	\mciteSetBstMidEndSepPunct{\mcitedefaultmidpunct}
	{\mcitedefaultendpunct}{\mcitedefaultseppunct}\relax
	\EndOfBibitem
	\bibitem[Mino \latin{et~al.}(2016)Mino, Kagawa, Ishigami, and
	Matsuyama]{mino2016}
	Mino,~Y.; Kagawa,~Y.; Ishigami,~T.; Matsuyama,~H. Numerical simulation of
	coalescence phenomena of oil-in-water emulsions permeating through straight
	membrane pore. \emph{Colloids Surf., A} \textbf{2016}, \emph{491}, 70\relax
	\mciteBstWouldAddEndPuncttrue
	\mciteSetBstMidEndSepPunct{\mcitedefaultmidpunct}
	{\mcitedefaultendpunct}{\mcitedefaultseppunct}\relax
	\EndOfBibitem
	\bibitem[Kagawa \latin{et~al.}(2014)Kagawa, Ishigami, Hayashi, Fuse, Mino, and
	Matsuyama]{kagawa2014}
	Kagawa,~Y.; Ishigami,~T.; Hayashi,~K.; Fuse,~H.; Mino,~Y.; Matsuyama,~H.
	Permeation of concentrated oil-in-water emulsions through a membrane pore:
	numerical simulation using a coupled level set and the volume-of-fluid
	method. \emph{Soft Matter} \textbf{2014}, \emph{10}, 7985\relax
	\mciteBstWouldAddEndPuncttrue
	\mciteSetBstMidEndSepPunct{\mcitedefaultmidpunct}
	{\mcitedefaultendpunct}{\mcitedefaultseppunct}\relax
	\EndOfBibitem
	\bibitem[Costa \latin{et~al.}(2017)Costa, Gomes, and Cunha]{costa2017studies}
	Costa,~A. L.~R.; Gomes,~A.; Cunha,~R.~L. Studies of droplets formation regime
	and actual flow rate of liquid-liquid flows in flow-focusing microfluidic
	devices. \emph{Exp. Therm. Fluid Sci.} \textbf{2017}, \emph{85}, 167\relax
	\mciteBstWouldAddEndPuncttrue
	\mciteSetBstMidEndSepPunct{\mcitedefaultmidpunct}
	{\mcitedefaultendpunct}{\mcitedefaultseppunct}\relax
	\EndOfBibitem
	\bibitem[Cao \latin{et~al.}(2018)Cao, Wu, and Sund{\'e}n]{cao2018dimensionless}
	Cao,~Z.; Wu,~Z.; Sund{\'e}n,~B. Dimensionless analysis on liquid-liquid flow
	patterns and scaling law on slug hydrodynamics in cross-junction
	microchannels. \emph{Chem. Eng. J.} \textbf{2018}, \emph{344}, 604\relax
	\mciteBstWouldAddEndPuncttrue
	\mciteSetBstMidEndSepPunct{\mcitedefaultmidpunct}
	{\mcitedefaultendpunct}{\mcitedefaultseppunct}\relax
	\EndOfBibitem
	\bibitem[Sussman \latin{et~al.}(1994)Sussman, Smereka, and Osher]{suss-1994}
	Sussman,~M.; Smereka,~P.; Osher,~S. A level set approach for computing
	solutions to incompressible two-phase flow. \emph{J. Comput. Phys.}
	\textbf{1994}, \emph{114}, 146\relax
	\mciteBstWouldAddEndPuncttrue
	\mciteSetBstMidEndSepPunct{\mcitedefaultmidpunct}
	{\mcitedefaultendpunct}{\mcitedefaultseppunct}\relax
	\EndOfBibitem
	\bibitem[Hirt and Nichols(1981)Hirt, and Nichols]{hirt-1981}
	Hirt,~C.~W.; Nichols,~B.~D. Volume of fluid (VOF) method for the dynamics of
	free boundaries. \emph{J. Comput. Phys.} \textbf{1981}, \emph{39}, 201\relax
	\mciteBstWouldAddEndPuncttrue
	\mciteSetBstMidEndSepPunct{\mcitedefaultmidpunct}
	{\mcitedefaultendpunct}{\mcitedefaultseppunct}\relax
	\EndOfBibitem
	\bibitem[Brackbill \latin{et~al.}(1992)Brackbill, Kothe, and
	Zemach]{brack-1992}
	Brackbill,~J.; Kothe,~D.~B.; Zemach,~C. A continuum method for modeling surface
	tension. \emph{J. Comput. Phys.} \textbf{1992}, \emph{100}, 335\relax
	\mciteBstWouldAddEndPuncttrue
	\mciteSetBstMidEndSepPunct{\mcitedefaultmidpunct}
	{\mcitedefaultendpunct}{\mcitedefaultseppunct}\relax
	\EndOfBibitem
	\bibitem[Sussman \latin{et~al.}(1999)Sussman, Almgren, Bell, Colella, Howell,
	and Welcome]{sussman1999}
	Sussman,~M.; Almgren,~A.~S.; Bell,~J.~B.; Colella,~P.; Howell,~L.~H.;
	Welcome,~M.~L. An adaptive level set approach for incompressible two-phase
	flows. \emph{J. Comput. Phys.} \textbf{1999}, \emph{148}, 81\relax
	\mciteBstWouldAddEndPuncttrue
	\mciteSetBstMidEndSepPunct{\mcitedefaultmidpunct}
	{\mcitedefaultendpunct}{\mcitedefaultseppunct}\relax
	\EndOfBibitem
	\bibitem[Issa(1986)]{issa1986}
	Issa,~R.~I. Solution of the implicitly discretised fluid flow equations by
	operator-splitting. \emph{J. Comput. Phys.} \textbf{1986}, \emph{62},
	40\relax
	\mciteBstWouldAddEndPuncttrue
	\mciteSetBstMidEndSepPunct{\mcitedefaultmidpunct}
	{\mcitedefaultendpunct}{\mcitedefaultseppunct}\relax
	\EndOfBibitem
	\bibitem[Holt(2012)]{holt2012}
	Holt,~M. \emph{\textit{Numerical methods in fluid dynamics}}; Springer Science
	\& Business Media, 2012\relax
	\mciteBstWouldAddEndPuncttrue
	\mciteSetBstMidEndSepPunct{\mcitedefaultmidpunct}
	{\mcitedefaultendpunct}{\mcitedefaultseppunct}\relax
	\EndOfBibitem
	\bibitem[Yao \latin{et~al.}(2018)Yao, Liu, Xu, Zhao, and Chen]{yao2017}
	Yao,~C.; Liu,~Y.; Xu,~C.; Zhao,~S.; Chen,~G. Formation of Liquid-Liquid Slug
	Flow in a Microfluidic T-junction: Effects of Fluid Properties and Leakage
	Flow. \emph{AIChE J.} \textbf{2018}, \emph{64}, 346\relax
	\mciteBstWouldAddEndPuncttrue
	\mciteSetBstMidEndSepPunct{\mcitedefaultmidpunct}
	{\mcitedefaultendpunct}{\mcitedefaultseppunct}\relax
	\EndOfBibitem
	\bibitem[Sontti and Atta(2017)Sontti, and Atta]{sontti2017cfdsa}
	Sontti,~S.~G.; Atta,~A. CFD Analysis of Taylor Bubble in a Co-Flow Microchannel
	with Newtonian and Non-Newtonian Liquid. \emph{Ind. Eng. Chem. Res.}
	\textbf{2017}, \emph{56}, 7401\relax
	\mciteBstWouldAddEndPuncttrue
	\mciteSetBstMidEndSepPunct{\mcitedefaultmidpunct}
	{\mcitedefaultendpunct}{\mcitedefaultseppunct}\relax
	\EndOfBibitem
	\bibitem[Li and Angeli(2017)Li, and Angeli]{li2017experimental}
	Li,~Q.; Angeli,~P. Experimental and numerical hydrodynamic studies of ionic
	liquid-aqueous plug flow in small channels. \emph{Chem. Eng. J.}
	\textbf{2017}, \emph{328}, 717\relax
	\mciteBstWouldAddEndPuncttrue
	\mciteSetBstMidEndSepPunct{\mcitedefaultmidpunct}
	{\mcitedefaultendpunct}{\mcitedefaultseppunct}\relax
	\EndOfBibitem
	\bibitem[Sontti and Atta(2019)Sontti, and Atta]{sontti2019cfdf}
	Sontti,~S.~G.; Atta,~A. CFD study on Taylor bubble characteristics in
	Carreau-Yasuda shear thinning liquids. \emph{Can. J. Chem.} \textbf{2019},
	\emph{97}, 616\relax
	\mciteBstWouldAddEndPuncttrue
	\mciteSetBstMidEndSepPunct{\mcitedefaultmidpunct}
	{\mcitedefaultendpunct}{\mcitedefaultseppunct}\relax
	\EndOfBibitem
	\bibitem[Sontti and Atta(2017)Sontti, and Atta]{sontti2017cfd}
	Sontti,~S.~G.; Atta,~A. CFD analysis of microfluidic droplet formation in
	non--Newtonian liquid. \emph{Chem. Eng. J.} \textbf{2017}, \emph{330},
	245\relax
	\mciteBstWouldAddEndPuncttrue
	\mciteSetBstMidEndSepPunct{\mcitedefaultmidpunct}
	{\mcitedefaultendpunct}{\mcitedefaultseppunct}\relax
	\EndOfBibitem
	\bibitem[Abiev(2017)]{abiev2017analysis}
	Abiev,~R.~S. Analysis of local pressure gradient inversion and form of bubbles
	in Taylor flow in microchannels. \emph{Chem. Eng. Sci.} \textbf{2017},
	\emph{174}, 403\relax
	\mciteBstWouldAddEndPuncttrue
	\mciteSetBstMidEndSepPunct{\mcitedefaultmidpunct}
	{\mcitedefaultendpunct}{\mcitedefaultseppunct}\relax
	\EndOfBibitem
\end{mcitethebibliography}
\providecommand{\latin}[1]{#1}
\makeatletter
\providecommand{\doi}
{\begingroup\let\do\@makeother\dospecials
	\catcode`\{=1 \catcode`\}=2 \doi@aux}
\providecommand{\doi@aux}[1]{\endgroup\texttt{#1}}
\makeatother
\providecommand*\mcitethebibliography{\thebibliography}
\csname @ifundefined\endcsname{endmcitethebibliography}
{\let\endmcitethebibliography\endthebibliography}{}

\newpage

\section*{For Table of Contents Only}
\begin{figure}
	\centering
	\includegraphics[width=\textwidth]{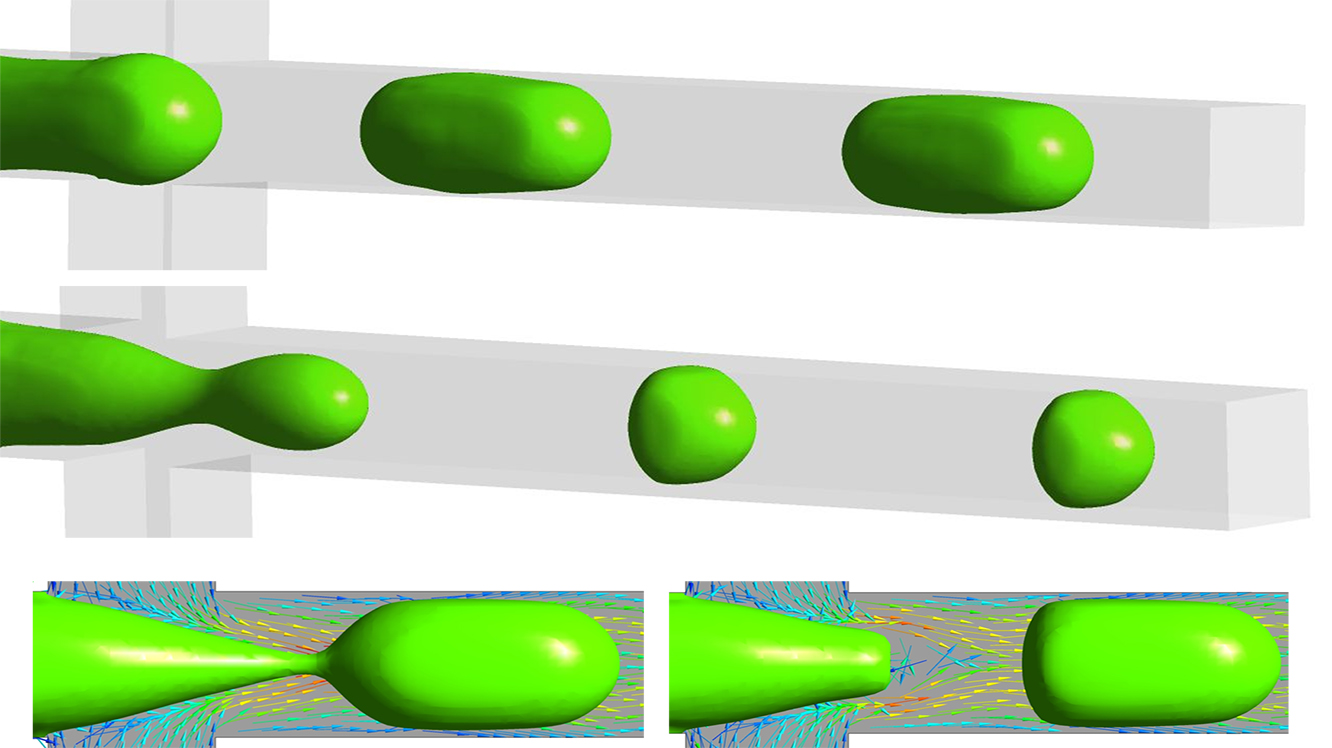}
	\caption*{\label{TOC} For Table of Contents Only}
\end{figure}

\end{document}